\documentclass[ALICE,manyauthors]{cernphprep}

\newcommand{\pp}{p\kern-0.05em p}
\newcommand{\ppbar}{\mathrm{p}\kern-0.05em \overline{\mathrm{p}}}

\newcommand{\GeV}{\ensuremath{\mathrm{GeV}\kern-0.05em}}
\newcommand{\GeVc}{\ensuremath{\mathrm{GeV}\kern-0.05em/\kern-0.02em c}}

\newcommand{\pT}{\ensuremath{p_{\mathrm{T}}}}

\newcommand{\pTjet}{\ensuremath{p_{\mathrm{T}}^{\mathrm{jet}}}}
\newcommand{\pTchjet}{\ensuremath{p_{\mathrm{T}}^{\mathrm{ch.\; jet}}}}

\usepackage[comma,square,numbers,sort&compress]{natbib}
\usepackage{hyperref}
\usepackage{lineno}
\usepackage{amsmath}
\usepackage{booktabs}
\usepackage{bigints}  
\usepackage{afterpage}  
\usepackage{multirow}
\usepackage{xcolor}
\usepackage[Symbol]{upgreek}
\usepackage[T1]{fontenc}
\usepackage{mathtools}
\usepackage{siunitx}
\usepackage{soul}
\usepackage{orcidlink}

\DeclarePairedDelimiterXPP\BigOSI[2]%
  {\mathcal{O}}{(}{)}{}%
  {\SI{#1}{#2}}

\usepackage{xspace}
\usepackage{color}
\usepackage{xcolor}
\usepackage{rotating}
\definecolor{light-gray}{gray}{0.8}

\newcommand{\MeVc}{\ensuremath{{\rm MeV/}c}\xspace}

\newcommand{\TeV}{\ensuremath{\rm TeV}\xspace}

\newcommand{\pt}{\ensuremath{{\it p}_{\rm T}}\xspace}

\newcommand{\ptDrange}[2]{{#1 $<$ \ptD $<$ #2  \GeVc}\xspace}
\newcommand{\etaless}[1]{\ensuremath{\left|\eta\right| < #1}\xspace}

\newcommand{\yless}[1]{\ensuremath{\left|y\right| < #1}\xspace}

\newcommand{\ptD}{\ensuremath{{\it p}_{\rm T}^{\rm D^{0}}}\xspace}

\newcommand{\Dzero}{\ensuremath{{\rm D}^{0}}\xspace}
\newcommand{\Dzerobar}{\ensuremath{\overline{{\rm D}^{0}}}\xspace}

\newcommand{\cmnt}[1]{}

\newcommand{\jetptrange}[2]{{#1 $<$ \pTchjet $<$ #2  \GeVc}\xspace}

\newcommand{\ptjet}{\ensuremath{{\it p}_{\rm T}^{\rm jet}}\xspace}

\begin{document}%

\begin{titlepage}
\PHyear{2026}
\PHnumber{170}      
\PHdate{06 June}   
%

\title{Probing flavor effects in the QCD parton shower using $\mathbf{\Dzero}$-tagged jet angularities in proton--proton collisions at $\mathbf{\sqrt{\textit{s}} = 5.02}$~TeV} 

\ShortTitle{}   

\Collaboration{ALICE Collaboration\thanks{See Appendix~\ref{app:collab} for the list of collaboration members}}
\ShortAuthor{ALICE Collaboration} 

\begin{abstract}
The ALICE Collaboration presents the first measurements of \Dzero-tagged jet angularities in proton--proton (pp) collisions at {\mbox{$\sqrt{s} = 5.02$~\TeV{}}}. Jet angularities are powerful substructure observables that characterize the angular and momentum distributions of particles within jets via tunable weighting parameters. Varying the angular parameter in jet angularities allows for a systematic probe of the sensitivity to collinear and soft radiation, enabling the study of flavor-dependent fragmentation and hadronization through comparisons of jets initiated by different partons. This paper reports \Dzero-tagged and inclusive (gluon-dominated) charged-particle jet angularities with a resolution parameter $R=0.4$ in the low jet transverse momentum range (\mbox{$10 < \pTchjet < 20$ \GeVc}), where charm-quark mass effects are most significant. At low angular weight, which emphasizes collinear radiation, \Dzero-tagged jets exhibit smaller angularity values than inclusive jets. This provides evidence for the radiation suppression from massive quarks -- a phenomenon known as the QCD dead-cone effect. As the angular weight increases, giving more emphasis to wide-angle radiation, the difference between \Dzero-tagged and inclusive jet distributions decreases. This indicates that the modification is concentrated within the jet core rather than its edge. PYTHIA 8 simulations qualitatively reproduce both the angularity of \Dzero-tagged and inclusive charged-particle jets, but reproduce the \Dzero-tagged jet distributions better than those of inclusive jets, offering a powerful new constraint for models. These results provide insight into flavor-dependent fragmentation and establish an essential baseline for future studies of jet modifications in the quark--gluon plasma produced in heavy-ion collisions.

\end{abstract}
\end{titlepage}
\setcounter{page}{2}

\section{Introduction}\label{sec:intro}
Quantum chromodynamics (QCD), the fundamental theory of the strong interaction, predicts that the gluon emission probability from a parton (quark or gluon) depends on both its mass and color charge~\cite{Altarelli:1977zs, BPQCD, Dokshitzer:1991fd}. These predictions are tested through detailed studies of high-energy collisions at particle accelerators, such as those at the CERN Large Hadron Collider (LHC). In these collisions, energetic partons scatter off one another with large momentum transfer ($Q^{2}$). In the parton shower picture, these hard-scattered partons dissipate energy through a cascade of gluon radiation and subsequent quark–antiquark pair production, forming a collimated spray of partons. As their excess mass decreases to the characteristic scale of QCD, $\Lambda_{\rm QCD}$, the partons hadronize, producing collimated sprays of observable hadrons known as jets. Since QCD predicts that the Altarelli–Parisi splitting probability for gluon radiation is directly proportional to the emitter’s color charge~\cite{Altarelli:1977zs}, gluon-initiated showers (carrying a larger color charge than quarks) are typically broader and produce more low transverse momentum (\pt{}) hadrons than quark-initiated showers. Heavy-quark-initiated (charm or beauty) showers are more collimated due to the large mass of the heavy quark, resulting in the dead-cone effect~\cite{Dokshitzer:1991fd}, which suppresses collinear gluon radiation and leads to harder fragmentation
~\cite{2023}.

Jets serve as powerful tools for probing the dynamics of QCD across a wide range of energy scales. This includes production of hard-scattered partons and the subsequent parton shower (both perturbative QCD calculable processes) as well as hadronization, a non-perturbative phenomenon that remains theoretically challenging to describe from first principles. Measurements of heavy-quark–initiated jets enable the disentanglement of parton-mass and
color-charge effects on both the parton shower and hadronization. Furthermore, the large mass of a heavy quark $(m_{\rm q} \gg \Lambda_{\rm QCD})$ introduces an additional hard scale to the scattering beyond $Q^{2}$. This hard scale effectively suppresses non-perturbative effects and ensures a stable perturbative expansion for heavy-flavor (HF) production cross sections.

Jets initiated by light quarks and gluons have been extensively studied~\cite{CMS:2017qlm, ATLAS:2019mgf, ALICE:2021obz, ALICE:EEC, ALICE:2022wpn, STAR:2026lyx}, providing important insights into the mechanisms of jet fragmentation and hadronization. More recently, several measurements of HF jets and their substructure have been reported by ALICE~\cite{ALICE:2019cbr, ALICE:2021wct, ALICE:deadcone, 2023, Acharya_2023, Acharya_2024, Acharya_2025, ALICE:DjetEEC}, ATLAS~\cite{Aad_2012, ATLAS:2013uet, Aad_2023}, CMS~\cite{CMS:2018ypj, Sirunyan:2019rfa, CMS:2019jis}, and LHCb~\cite{LHCb:2017llq, LHCb-PAPER-2020-018, Aaij_2022, lhcb_psi2s_chic13872, lhcb_bjet_mass, lhcb_bjet_LJP}, highlighting the growing interest in understanding the flavor dependence of jet formation and hadronization. Nevertheless, the roles of parton mass and color charge in these processes remain not well understood. This paper explores flavor-dependent parton shower evolution and hadronization using a class of jet-substructure observables known as generalized jet angularities~\cite{Larkoski_2014, Almeida_2009, Reichelt_2022}. Originally proposed as event-shape observables, generalized jet angularities probe both the \pt and angular distributions of constituents within the jet, with
contributions weighted by the parameters $\kappa$ and $\alpha$, respectively:
\begin{equation}
    \lambda_{\alpha}^{\kappa} \equiv { \sum_{i \in \rm jet} }\bigg(\frac{p_{{\rm T},i}}{p_{\rm T}^{\rm jet}}\bigg)^{\kappa}
\bigg(\frac{\Delta R_{{\rm jet},i}}{R}\bigg)^{\alpha} \equiv { \sum_{i \in \rm jet}} z_{i}^{\kappa}\theta_{i}^{\alpha},
 \label{eq:angularity}
\end{equation}
where the sum runs over all constituents $i$ within the jet, $R$ is the jet resolution parameter, and 
\begin{equation}
    \Delta R_\mathrm{jet,i} = \sqrt{(y_{\rm jet} - y_{\rm i})^{2} + (\varphi_{\rm jet} - \varphi_{\rm i})^{2}}
\end{equation}
is the distance between the constituent and the jet axis in the rapidity--azimuthal angle ($y-\varphi$) plane. These observables are infrared and collinear safe (IRC safe)\footnote{Track-based observables are IRC-unsafe; however, comparisons to theoretical predictions can
nonetheless be carried out by following a non-perturbative correction procedure. See Ref.~\cite{Alice_ang_2022} for details.} for configurations with $\kappa = 1$ and {\mbox{$\alpha > 0$}}~\cite{Berger_2003, Kang_2018}, ensuring theoretical calculability.

The value of $\alpha$ can be tuned to probe jet substructure across a range of angular scales where smaller values of $\alpha$ emphasize contributions from collinear radiation, and larger values enhance sensitivity to wide-angle radiation, enabling a detailed study of the jet angular profile. The CMS Collaboration previously measured generalized jet angularities for beauty-tagged jets~\cite{CMS:2018ypj}; however, the study indicated no significant differences between quark- and gluon-enriched samples for $\pTjet > 30~\GeVc$. This paper explores an extended kinematic coverage at lower \pTjet, where mass effects are expected to be more
significant.

Jet angularities have previously been studied by ALICE for inclusive jets with $\ptjet>20$ \GeVc~\cite{Alice_ang_2022}. Using the ALICE detector's excellent tracking and particle identification capabilities, jets can be tagged with fully reconstructed HF hadrons, resulting in an enriched sample of charm-initiated jets. This paper reports the first measurement of jet angularities for \Dzero-tagged jets in pp collisions at $\sqrt{s} = 5.02$ \TeV, where tagging is performed via the presence of a reconstructed \Dzero meson among the jet constituents. The measurement is performed for angularity parameter values $\alpha = 1$, 1.5, 2, and 3, and $\kappa = 1$. Since $\kappa$ is fixed to unity, the observable is denoted as $\lambda_{\alpha}$.
In addition, jet angularities are measured for inclusive jets in the same kinematic region as the \Dzero-tagged jets, both with and without a leading-track \pt selection within the jets. Both \Dzero-tagged and inclusive jets are reconstructed as charged-particle jets (also referred to as charged jets). Furthermore, these measurements are compared with state-of-the-art simulations, providing new constraints on Monte Carlo (MC) event generators and hadronization models. Finally, these results provide an essential baseline for future studies in heavy-ion collisions.
 
\section{Experimental setup and data sets}
\label{sec:exp_data}
Heavy-flavor hadrons and jets are reconstructed using the central barrel detectors of the ALICE experiment. The primary detectors are the Inner Tracking System (ITS)~\cite{CERN-LHCC-99-012}, the Time Projection Chamber (TPC)~\cite{Dellacasa:451098}, and the Time-of-Flight (TOF) detector~\cite{CERN-LHCC-2000-012,Cortese:545834}. They are embedded in a large solenoidal magnet that provides a 0.5 T magnetic field parallel to the beam axis. The central barrel ensures a high reconstruction efficiency for charged particles in $0.15 < \pT <100 \ \GeVc$, with uniform acceptance in $|\eta|<0.9$ and full azimuth $(0 < \varphi < 2\uppi)$. The ITS is the innermost tracking detector of ALICE, designed to provide high-resolution tracking and vertex reconstruction. It consists of six cylindrical layers of silicon detectors: two layers of each of the following-- Silicon Pixel Detectors, Silicon Drift Detectors, and Silicon Strip Detectors. The ITS plays a crucial role in the precise determination of primary and secondary vertices, as well as in charged-particle tracking at low $p_{\rm T}$.

The TPC is ALICE's primary tracking detector and is crucial for identifying charged particles. It consists of a large cylindrical gas-filled chamber with readout planes at either end. Charged particles traversing the TPC ionize the gas, and the resulting electrons drift toward the readout planes under an electric field. The TPC provides excellent momentum resolution and particle identification via specific energy-loss (d$E$/d$x$) measurements. The TOF detector is a large array of Multi-gap Resistive Plate Chambers (MRPCs) located outside the TPC. It is used for particle identification by measuring the time particles take to travel from the interaction point to the TOF, with a precision better than a tenth of a billionth of a second. The TOF provides complementary particle identification to the TPC, enabling the separation of pions, kaons, and protons in the intermediate-$p_{\rm T}$ range based on their mass-dependent velocity differences. A complete description of the ALICE detector and an overview of its performance are presented in Refs.~\cite{ALICE:2022wpn,Aamodt:2008zz, Abelev:2014ffa}. 

The results presented in this paper are obtained using data recorded by ALICE during the LHC Run 2 in 2017 for pp collisions at a center-of-mass energy of $\sqrt{s}=5.02$ TeV. Events used in this analysis are selected using the minimum-bias (MB) trigger, defined by a coincident signal in the V0 scintillator detectors~\cite{ALICE:2013axi}. The V0 detector consists of two scintillator arrays placed on either side of the interaction point along the beam axis, covering the pseudorapidity ranges $2.8 < \eta < 5.1$ (V0A) and $-3.7 < \eta < -1.7$ (V0C). To ensure uniform detector acceptance, events are required to have a primary vertex located within $\pm$10 cm of the nominal interaction point. Events with more than one reconstructed vertex are excluded to avoid pileup effects. After applying these selection criteria, a total of 870 million MB events are retained, corresponding to an integrated luminosity of $(18.0\pm0.4)$ nb$^{-1}$~\cite{ppXsec}.

The Monte Carlo samples used for the corrections are generated using the PYTHIA 6.4.25 event generator~\cite{Sj_strand_2006} with the Perugia 2011 tune~\cite{Skands_2010}, and with the GEANT 3.21.11~\cite{Brun:118715} transport model. The ALICE detector layout, along with variations in the data-taking conditions during the run, is reproduced in the simulation. These samples are referred to as PYTHIA 6 + GEANT 3 in the subsequent discussion, unless otherwise specified.
\section{Analysis method}
\label{sec:analysis}
In this measurement, heavy-quark jets originating from charm quarks are identified by the presence of a prompt \Dzero meson among their constituents. Prompt \Dzero mesons are produced directly from the charm-quark fragmentation, whereas those originating from the decays of beauty hadrons are referred to as non-prompt \Dzero mesons. Prior to jet finding, D$^0$-meson candidates are reconstructed via the decay mode ${\rm D^{0} \rightarrow K^{-} \pi^{+}}$  (with branching ratio \mbox {{${\rm 3.947} \pm {\rm 0.030}\%$}}~\cite{pdg2024}) within the rapidity range \yless0.8, selecting \Dzero mesons in the \ptDrange{5}{36} interval. The \Dzero meson and its antiparticle (\Dzerobar) are treated equivalently and are referred to collectively as \Dzero. The final corrected distributions of $\lambda_{\alpha}$ for \Dzero-tagged jets (procedure detailed below) are shown for jets in the interval \jetptrange{10}{20}, requiring a \Dzero meson in the interval \ptDrange{5}{20}.

\subsection{\texorpdfstring{${\mathbf{D^{0}}}$}{D0}-meson and jet reconstruction}
The decay-vertex topology of the \Dzero-meson candidates is fully reconstructed by combining two charged tracks with the appropriate charge sign, $\pt>0.3$~\GeVc, and $|\eta|<0.8$. Tracks must have at least 70 associated TPC space points, $\chi^{2}/{\rm ndf}<2$ in the TPC (where ndf is the number of degrees of freedom involved in the track fit procedure), and at least one hit in either of the two layers of the SPD. Particle identification of the \Dzero-meson daughters uses specific energy loss (d$E$/d$x$) measured in the TPC and the time-of-flight information provided by the TOF detector. Pions and kaons are selected within $3 \sigma$ of the expected mean values, where $\sigma$ represents the resolution of the d$E$/d$x$ and time-of-flight measurements. Particles without TOF information were identified using only the TPC data. \Dzero-meson selection out of the large combinatorial background is achieved by exploiting topological selections on variables such as the distance of closest approach (DCA) between the two daughter tracks, the \pt of daughter tracks, the impact parameters of the daughter particles and their product, and the cosine of the \Dzero-meson decay angle. Further details about the track and \Dzero-meson candidates selections can be found in Ref~\cite{ALICE:2019cbr}. A small fraction of cases have inconclusive PID and result in signal candidates passing the selection as both \Dzero and \Dzerobar, with swapped mass assignments for the decay particles. These are referred to as a signal reflection. The contribution from these is removed following the strategy explained in the next section. 

Using the aforementioned topological selection cuts, a significant fraction of the combinatorial background from uncorrelated daughter tracks is removed, providing a high \Dzero signal-to-background ratio. The decay daughters of the \Dzero-meson candidates are replaced by their four-momentum vector sum (i.e. the reconstructed \Dzero meson). \Dzero-tagged jets are reconstructed from \Dzero candidates and charged-particle tracks using the anti-$k_\mathrm{T}$ algorithm~\cite{Cacciari_2008} implemented in FastJet~\cite{Cacciari:2011ma}, with resolution parameter $R = 0.4$ and $E$-scheme recombination. For reconstructed jets, the $m_{\mathrm{D}^0}$ (mass of \Dzero at detector level) is assigned to the \Dzero candidates, while all other tracks are assumed to be charged $\pi$ mesons with $m_{\mathrm{\pi}^\pm} = 0.1396$ \GeVc$^{2}$. The jet axis is required to have \etaless 0.5 to ensure that the full jet cone is within the TPC acceptance. A minimum \pt cut of 150 \MeVc~ is applied to tracks and a $p^{\rm ch.\ jet}_{\rm T}$ cut of 5 \GeVc~is imposed on reconstructed jets. Jets containing a \Dzero candidate among their constituents are retained for further analysis. 

\subsection{Raw yield extraction}\label{sec:rawyield}
After jet reconstruction, the angularity ($\lambda_{\alpha}$) is calculated for each jet. At this stage, the angularity distribution includes contributions from jets containing a real \Dzero meson, as well as jets containing background \Dzero candidates arising from random combinations of pions and kaons that do not originate from a true \Dzero meson, in addition to contributions from signal reflections. 

The background is removed using an invariant-mass ($M$) sideband subtraction procedure, as explained below. The amount of background \Dzero candidates that pass the selection criteria depends on their transverse momentum (\ptD); therefore, the subtraction is done in intervals of \ptD as well as in intervals of jet \pt. Consequently, a two-dimensional distribution of $\lambda_{\alpha}$ versus $M$ for \Dzero-tagged jets is constructed and binned in candidate \ptD intervals within \ptjet intervals. The intervals depend on the statistical precision, fit quality, signal-to-background ratio, and signal significance. The signal significance and the signal-to-background ratio vary from 4 to 14 and from 2 to 5, respectively, depending on the \ptD interval.

The invariant-mass distributions are fitted with a Gaussian function to describe the \Dzero peak and an exponential function to describe the background. Reflection templates are obtained from simulations using the PYTHIA 6 event generator and are parametrized as a sum of two Gaussians, with the means, widths, and the \Dzero signal-to-reflection ratio fixed to values determined from simulations.  

\begin{figure}[!ht]
    \centering
     \includegraphics[width=0.45\textwidth]{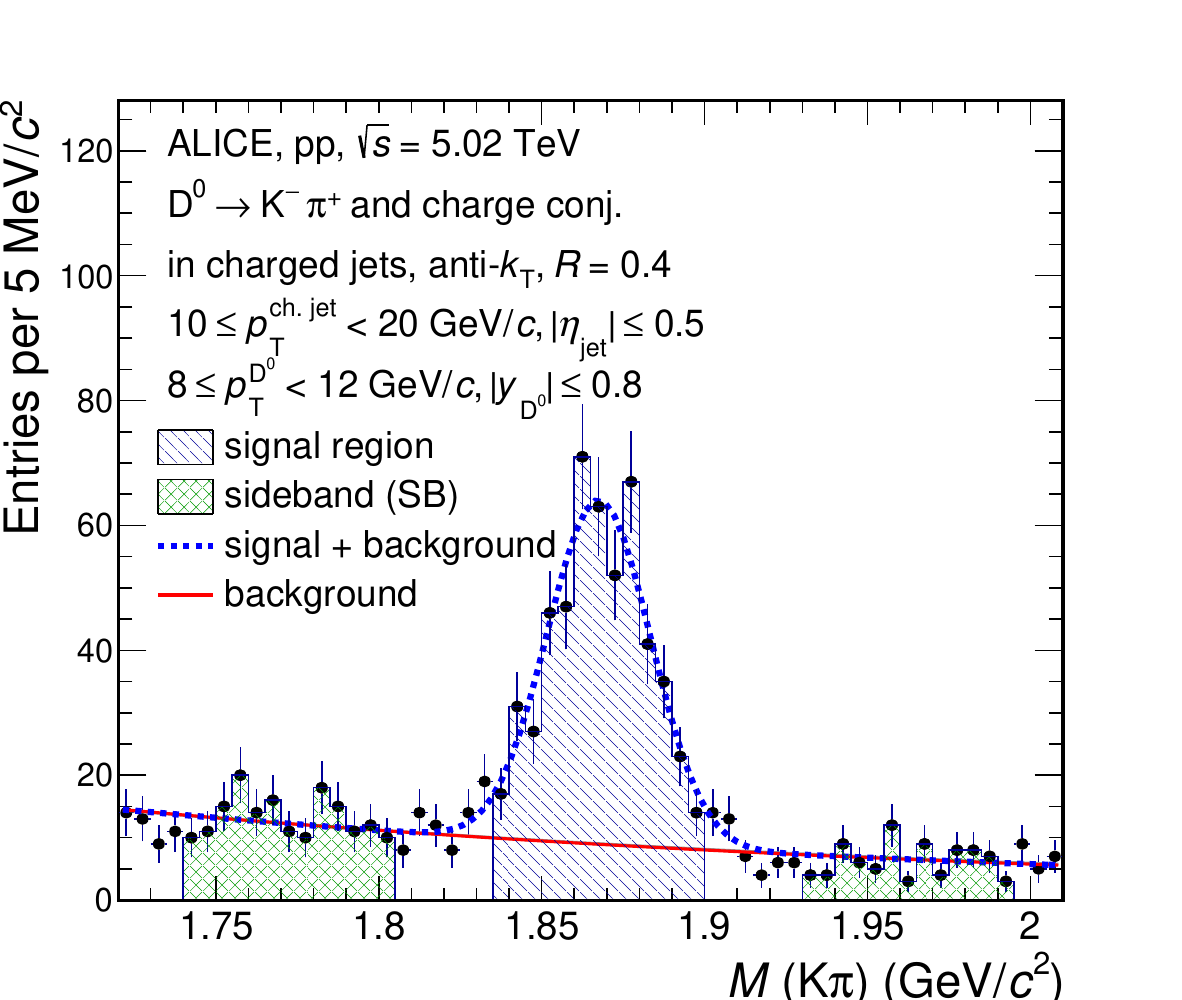}
     \includegraphics[width=0.45\textwidth]{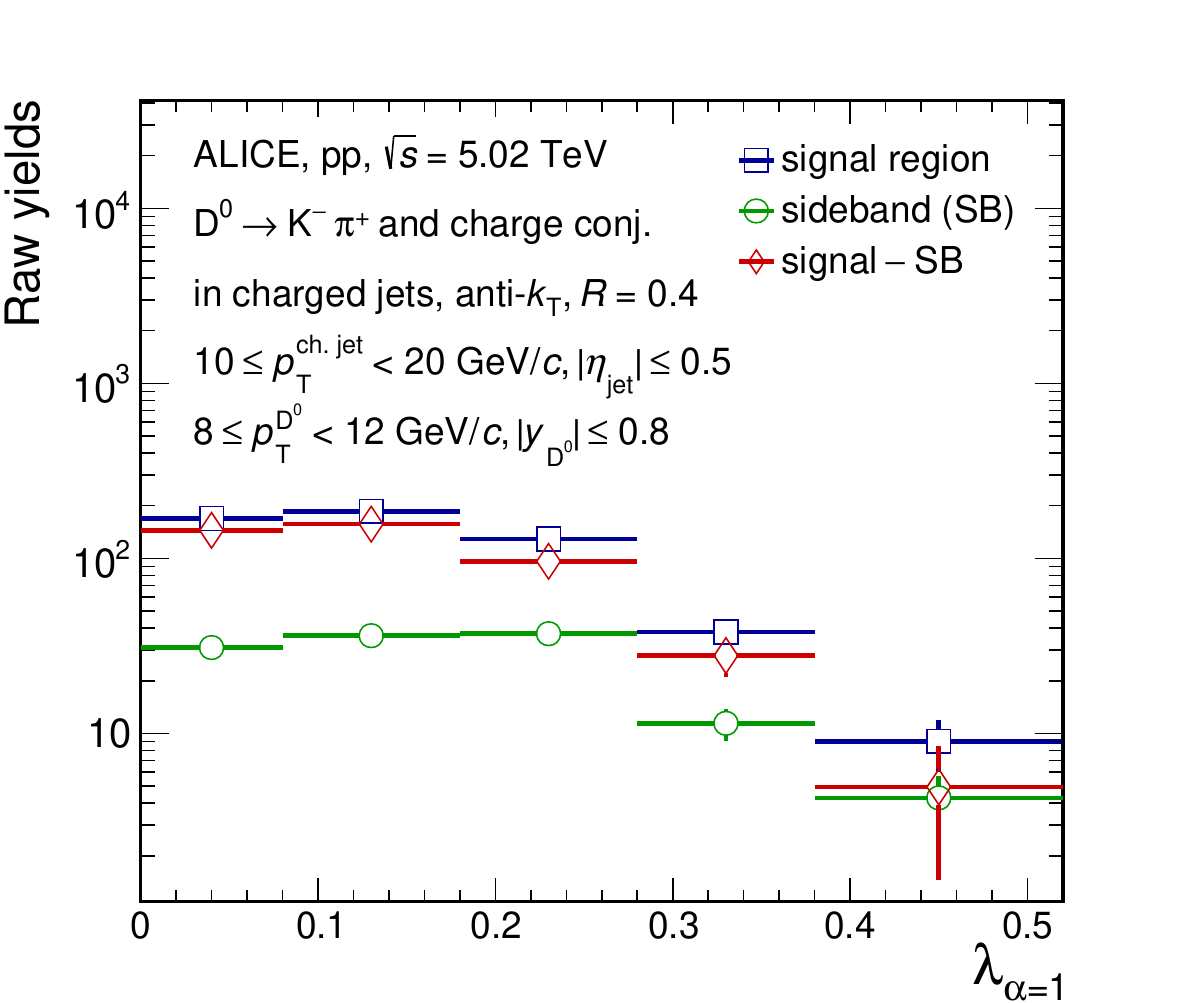}
   
    \caption{Left: Invariant-mass distribution of \Dzero-jet candidates for \jetptrange{10}{20}, with \Dzero-meson candidates in \ptDrange{8}{12} and $R=0.4$.
    The total fit function is depicted by the dashed blue line, while the solid red line represents the background fit function. The blue and green shaded areas correspond to the signal and sideband regions, respectively. Right: The distribution of the raw \Dzero-tagged jet yields as a function of jet angularity $\lambda_{\alpha=1}$ in the signal (blue) and sideband (green) regions, along with the background-subtracted (red) yields.}
    \label{fig:invariantmass}
\end{figure}

Figure \ref{fig:invariantmass} shows an example of the invariant-mass distribution (left panel) and the distribution of the raw \Dzero-tagged jet yields as a function of $\lambda_{\alpha=1}$ (right panel), for jets in the \jetptrange{10}{20} interval and \Dzero candidates in the {\mbox{\ptDrange{8}{12}}} interval. The signal region is defined as $|{ M - \mu_{\rm fit}}| < 2\sigma_{\rm fit}$, where $\mu_{\rm fit}$ is the mean, and $\sigma_{\rm fit}$ is the width of the Gaussian-fit component. The background regions (left and right sidebands) are chosen based on the condition, $4\sigma_{\rm fit} < |{ M \rm - \mu_{fit}}| < 9\sigma_{\rm fit}$. The signal and sideband regions are represented by the dashed blue and green areas, respectively. The reflection contributions are added to the background fit function. The right panel presents the raw yields of \Dzero-tagged jets as a function of $\lambda_{\alpha =1}$  extracted for the signal (blue) and sideband (green) invariant-mass regions. The sideband distributions, which characterize the background shape, are obtained by summing the left and right sidebands and are normalized to estimate the background contribution in the signal-peak region. The normalization is determined from the fitted invariant-mass background distribution and ensures that the sideband contribution is scaled to match the expected background under the signal peak region. This background is then subtracted from the signal-region distributions to obtain the distribution of the raw \Dzero-tagged jet-angularity $\lambda_{\alpha=1}$  distributions $(N_{\rm raw}(\lambda_{\alpha},p^{\rm ch.\ jet}_{\rm T,det}, \ptD))$. The combinatorial-background subtracted distribution is shown in red in the right panel of Fig.~\ref{fig:invariantmass}. After combinatorial-background corrections, a residual contribution from \Dzero mesons originating from beauty-hadron decays remains in the sample; the treatment of this feed-down component is described in the next section.

\subsection{Corrections}\label{sec:correction}
The raw $\lambda_{\alpha}$ distributions are corrected for three main factors: (i) the efficiency and acceptance of the \Dzero-tagged jet reconstruction, (ii) the subtraction of non-prompt \Dzero mesons (feed-down), and (iii) the momentum smearing introduced by detector effects.

\subsubsection{Reconstruction efficiency}
The distribution of raw \Dzero-tagged jet yields is corrected for the reconstruction and selection efficiencies ($\mathrm{Acc} \times \epsilon$). The efficiency is evaluated separately for prompt and non-prompt \Dzero-tagged jets using MC simulations, by taking the ratio of reconstructed \Dzero-tagged jets which pass all selection requirements to the total number of generated \Dzero-tagged jets within the TPC acceptance ($|\eta_{\rm jet}| < 0.5$). The reconstruction efficiency exhibits a strong dependence on \ptD~\cite{2023}, since the selection criteria are more stringent at low \ptD in order to suppress the combinatorial background. The prompt \Dzero-tagged jet efficiency varies between 2\%--30\%, while the non-prompt \Dzero-tagged jet efficiency ranges between 5\%--20\%. The signal extracted in intervals of \ptD and jet \pt is corrected with the \ptD-dependent efficiency and then summed over the whole \ptD range to obtain the yield in a given jet-\pt interval. The efficiency-corrected $N(\lambda_{\alpha},p^{\rm ch.\ jet}_{\rm T})$ distributions are obtained by the following equation:
\begin{equation}
    N(\lambda_{\alpha},p^{\rm ch.\ jet}_{\rm T})= \sum_{\ptD} \frac{N_{\rm raw}(\lambda_{\alpha},p^{\rm ch.\ jet}_{\rm T}, \ptD) }{({\rm Acc} \times \varepsilon)^{\rm c\rightarrow
 \Dzero}\ (\ptD)}.
 \label{eq:efficiency_corrected_n}
\end{equation}

\subsubsection{Subtraction of non-prompt \texorpdfstring{${\mathbf{\textit D^{0}}}$}{D0} contribution}
To study charm-quark fragmentation, the contribution to the raw yields due to non-prompt \Dzero mesons coming from beauty-hadron decays must be removed. The limited sample size precludes a data-driven estimation of the non-prompt \Dzero-tagged jet fraction. Instead, next-to-leading order (NLO) pQCD calculations from POWHEG~\cite{Alioli2010}, coupled with the PYTHIA 8 Monash tune~\cite{Skands:2014pea} parton shower, are used to estimate this contribution. The procedure to remove it is described below.

The POWHEG + PYTHIA 8 non-prompt \Dzero-tagged jet $\lambda_{\alpha}$ distributions are first multiplied by the non-prompt \Dzero-tagged jet efficiency to obtain the expected non-prompt \Dzero-tagged jets in data at the detector level. The resulting $\lambda_{\alpha}$ values are then scaled by the prompt \Dzero-tagged jet efficiency to match the efficiency-corrected distributions of non-prompt \Dzero-tagged jets in data. This ensures that the non-prompt \Dzero contribution is treated consistently with the prompt component. 

In the final step, the $\lambda_{\alpha}$ distributions are smeared by multiplying them with a response matrix $( {RM})$ for non-prompt-\Dzero-tagged jets ($RM^{\mathrm{b}\rightarrow \Dzero}_{\rm det}(\lambda_{\alpha}^{\rm det},\lambda_{\alpha}^{\rm truth},p^{\rm ch.\ jet}_{\rm T,det},p^{\rm ch.\ jet}_{\rm T,truth})$). The $RM$ encodes the correspondence between the true (particle-level) and measured (detector-level) distributions, thereby mapping the non-prompt \Dzero-tagged jet truth-level variables from PYTHIA 6 to  the detector-level variables reconstructed in the full PYTHIA 6 + GEANT 3 detector simulations.  The $ {RM}$ is re-weighted by the inverse of the prompt \Dzero-tagged jet efficiency to account for the fact that the measured sample is already corrected for it. Additional corrections are applied to account for jets that exist at the detector level but not at the truth level (fake jets), as well as jets that exist at the truth level but not at the detector level (missed jets)~\cite{2023}. The non-prompt \Dzero-meson-subtracted $\lambda_{\alpha}$ distributions are obtained using the following equation
\begin{equation}
 N^{ \rm c \rightarrow D^{0}}(\lambda_{\alpha},p^{\rm ch.\ jet}_{\rm T,det})  = N(\lambda_{\alpha},p^{\rm ch.\ jet}_{\rm T})  
    - N^{ \rm b \rightarrow D^{0}}(\lambda_{\alpha},p^{\rm ch.\ jet}_{\rm T,det})
\label{eq:feeddowncorrection3}
\end{equation}
where:
{\begin{itemize}
    \item[-]c and b stand for charm (prompt \Dzero) and beauty (non-prompt \Dzero), respectively;
    \item[-]$N^{ \rm c \rightarrow D^{0}}(\lambda_{\alpha},p^{\rm ch.\ jet}_{\rm T,det})$ denotes the efficiency-corrected measured $\lambda_{\alpha}$ distributions after subtraction of the non-prompt \Dzero contribution;
    \item[-]$N(\lambda_{\alpha},p^{\rm ch.\ jet}_{\rm T})$ (from Eq.~\ref{eq:efficiency_corrected_n}) is the total efficiency-corrected measured $\lambda_{\alpha}$ distributions, before subtraction of the b-jet contribution;
     \item[-] $N^{\mathrm{b}\rightarrow \Dzero}(\lambda_{\alpha},p^{\rm ch.\ jet}_{\rm T,det})$ is the non-prompt \Dzero-tagged jet $\lambda_{\alpha}$ distribution at detector level, given by the following equation
\end{itemize}}
\begin{equation}
 \begin{split}
     N^{\rm b \rightarrow D^{0}}(\lambda_{\alpha}^{\rm det},p^{\rm ch.\ jet}_{\rm T,det}) &= RM^{\rm b \rightarrow D^{0}}_{\rm det}(\lambda_{\alpha}^{\rm det},\lambda_{\alpha}^{\rm truth},p^{\rm ch.\ jet}_{\rm T,det},p^{\rm ch.\ jet}_{\rm T,truth})\ 
     \\
     &\otimes \sum_{\ptD} \frac{({\rm Acc} \times \varepsilon)^{\rm b\rightarrow 
 \Dzero}(\ptD)}{({\rm Acc} \times \varepsilon)^{\rm c\rightarrow 
 \Dzero}(\ptD)} N_{\rm POWHEG}^{\rm b \rightarrow D^{0}}(\lambda_{\alpha}^{\rm truth},p^{\rm ch.\ jet}_{\rm T,truth},\ptD )
     \end{split}
     \label{eq:feeddowncorrection4}
\end{equation}
where: 
{\begin{itemize}
    \item[-]the symbol $\otimes$ should be interpreted as the convolution of the non-prompt $RM^{\mathrm{b}\rightarrow \Dzero}$ and the vector of the yields;
    \item[-]$({\rm Acc} \times \varepsilon)^{\rm c\rightarrow 
 \Dzero}(\ptD)$ and $({\rm Acc} \times \varepsilon)^{\rm b\rightarrow \Dzero}(\ptD)$ are the \ptD -dependent products of the reconstruction and selection efficiencies for prompt and non-prompt \Dzero-tagged jets, respectively;
 \item[-] $ N_{\rm POWHEG}^{\mathrm {b \rightarrow D^{0}}}(\lambda_{\alpha}^{\rm truth},p^{\rm ch.\ jet}_{\rm T,truth}, \ptD)$ is the non-prompt \Dzero-tagged jet $\lambda_{\alpha}$ 
 distribution from simulation.
\end{itemize}}

\begin{figure}[!ht]
\centering
  \includegraphics[width=0.5\textwidth]{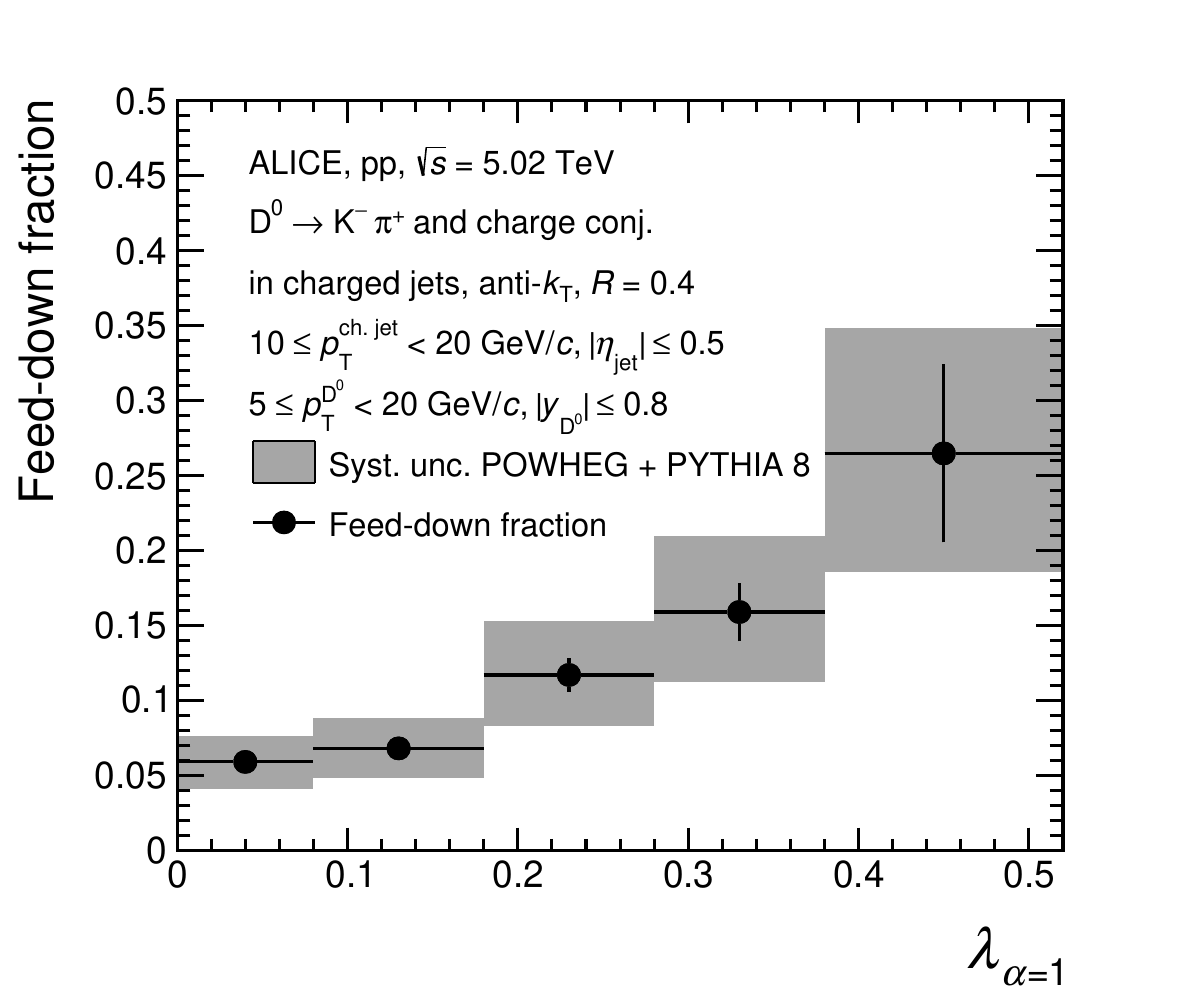}
\caption[]{Non-prompt (feed-down) fraction of \Dzero-tagged jets from beauty-hadron decays as a function of $\lambda_{\alpha=1}$ for jets with $R = 0.4$ and \jetptrange{10}{20}.}	
\label{fig:reconstructionefficiency_feeddown}
\end{figure}

Figure~\ref{fig:reconstructionefficiency_feeddown} shows the ratio of simulated non-prompt \Dzero-tagged jets to the efficiency-corrected total (prompt + non-prompt) \Dzero-tagged jets in data, as a function of $\lambda_{\alpha = 1}$ at detector level. The non-prompt fraction varies between 5\% and 25\% for all values of $\alpha$.

\subsubsection{Unfolding}
The $\lambda_{\alpha}$ distributions are finally corrected for detector effects, including track-\pt resolution, tracking inefficiencies, and particle interactions with the detector material. This correction is performed using a 2D unfolding in $p^{\rm ch.\ jet}_{\rm T}$ and $\lambda_{\alpha}$, employing an iterative Bayesian approach \cite{bayesianUnfolding}, as implemented in the RooUnfold package \cite{RooUnfold}. The 4D prompt response matrix, ${RM}^{\rm c \rightarrow D^{0}}(\lambda_{\alpha}^{\rm det},\lambda_{\alpha}^{\rm truth},p^{\rm ch.\ jet}_{\rm T,det},p^{\rm ch.\ jet}_{\rm T,truth})$, maps detector-level variables $(\lambda_{\alpha}^{\rm det}, p^{\rm ch.\ jet}_{\rm T,det})$, reconstructed in full PYTHIA 6 + GEANT 3 detector simulations, to the corresponding truth-level variables $(\lambda_{\alpha}^{\rm truth}, p^{\rm ch.\ jet}_{\rm T,truth})$. Before unfolding, the $RM$ is scaled by the prompt \Dzero-tagged jet efficiency to ensure consistency in the correction. The $\lambda_{\alpha}$ distributions are also corrected to account for fake jets and missed jets, similarly to the non-prompt \Dzero-tagged jet simulations.

To achieve convergence within 5\% between subsequent iterations while maintaining minimal statistical uncertainties, five unfolding iterations are performed for all $\alpha$ values. The stability of the unfolding procedure is validated through a series of closure tests, discussed in detail in Ref. \cite{Alice_ang_2022}. 

\subsection{Inclusive jet baseline measurement}
The inclusive-jet baseline measurement largely follows the same procedure as the \Dzero-tagged jet reconstruction (excluding steps specific to \Dzero reconstruction) and the published inclusive-jet angularity measurement in pp collisions at $\sqrt{s} = 5.02$ TeV ~\cite{Alice_ang_2022}. Inclusive jets are reconstructed from charged tracks with  $\pt > 150$ \MeVc using the anti-$k_{\rm T}$ algorithm (with the $E$ recombination scheme) in the interval \jetptrange{10}{20} for a resolution parameter of $R=0.4$. A subset of the inclusive jets is also constructed, with a leading-track \pt selection of $\pt > 5.33$ \GeVc~\cite{Acharya_2023}. This mimics the $\pt$ selection imposed on the \Dzero in the charm  jets, with the value of $5.33~\GeVc$ corresponding to the \pt of a $\pi^\pm$ meson with the same transverse mass as a \Dzero meson with $\pt = 5~\GeVc$. Inclusive jets are corrected for detector effects using the same procedure as for \Dzero-tagged jets.

\section{Systematic uncertainties}
\label{sec:systematics}
Systematic uncertainties in the \Dzero-tagged jet-angularity measurement arise as follows:

\begin{enumerate}
    \item \textbf{${\mathbf{D^{0}}}$-meson selection}: The systematic uncertainties associated with \Dzero-meson selection are evaluated by varying the topological and particle-identification criteria used. A combination of tighter and looser selection criteria are applied, resulting in variations of $\pm$10\% and $\pm$15\% in the reconstruction efficiency, respectively, relative to the default case. The final systematic uncertainty is assigned using the root mean square (RMS) of the distribution of differences between the central values of the angularity distributions and those obtained from the variations.
    \item \textbf{Signal extraction}: The uncertainty in the signal extraction (described in Sec. \ref{sec:rawyield}) is assessed by performing multiple fits to the invariant-mass distributions. Variations include: (i) changing the background function from an exponential (default) to a linear or polynomial function, (ii) modifying the fit range, (iii) adjusting the binning of the invariant mass distribution, (iv) varying the widths of the signal and background bands, and (v) treating the width $\sigma_{\rm fit}$ and $\mu_{\rm fit}$ of the Gaussian signal as either free parameters or fixing them to the values extracted from MC simulations. The systematic uncertainty is  assigned using the RMS of the distribution of the differences between the central values of the angularity distributions and those obtained from the variations. 
    \item \textbf{Non-prompt ${\mathbf{D^{0}}}$-meson correction}: The systematic uncertainty associated with the non-prompt \Dzero correction is estimated by varying key input parameters in the POWHEG + PYTHIA 8 simulations. These parameters include the b-quark mass, the renormalization and factorization scales in the perturbative calculation, and the choice of parton distribution function~\cite{Cacciari_2012}. The systematic uncertainty is taken as the maximum deviation from unity of the ratio of values obtained from variations to the default values.
    
    \item \textbf{Tracking efficiency}: The track-reconstruction-efficiency uncertainty is approximately 3\% in pp collisions, determined as a combination of uncertainties due to the TPC track-selection efficiency and the TPC–ITS matching efficiency~\cite{Acharya_2025}. To assess its impact, 3\% of the detector-level tracks in the MC simulation are randomly removed, and the analysis is repeated using the updated reconstruction efficiency and response matrix. The ratio of the variation to the default \Dzero-tagged jet angularity measurement is assigned as the corresponding systematic uncertainty.

    \item \textbf{Unfolding procedure}: To estimate the uncertainty associated with the unfolding procedure, four sources are considered: (i) The unfolding is repeated with a number of iterations varied by $\pm2$ around the central value (5 iterations), and the average difference from the nominal result is taken as the iteration uncertainty, (ii) the prior distribution is modified by applying a power-law scaling in $p_\mathrm{T}^\mathrm{ch.\;jet}$ and a linear scaling in $\lambda_{\alpha}$, $ (p_\mathrm{T}^\mathrm{ch.\;jet})^{\pm 0.5} \times (1 \pm 0.5 * (2 \lambda_\alpha - 1))$. The maximum absolute deviation from the nominal result is assigned as the prior uncertainty in each bin, (iii) the analysis is repeated using slightly finer and coarser bins than the default, and (iv) the lower $p_{\text{T, det}}^{\text{ch. jet}}$ limit is truncated by 1 GeV/$c$. The resulting deviation from the default \Dzero-tagged jet angularity measurement is assigned as the corresponding systematic uncertainty.
    
    These four sources probe the same source of uncertainty, so the total unfolding uncertainty is estimated here as the standard deviation of all four contributions.
\end{enumerate}

The total systematic uncertainty is calculated by assuming that the individual sources are uncorrelated and summing them in quadrature. A summary of the systematic uncertainties is provided in {\mbox{Tab.~\ref{tab:uncertainties_d0}.}}

\begin{table}[h]
    \centering
    \renewcommand{\arraystretch}{1.2}
    \caption{Systematic uncertainties of the \Dzero-tagged jet-angularities measurement for different $\alpha$ values. The minimum and maximum values of the uncertainties across the $\lambda_{\alpha}$ intervals are provided.}
    \begin{tabular}{ccccccc}
        \toprule
        \multicolumn{1}{c}{} & \multicolumn{6}{c}{Relative uncertainty} \\
        \cmidrule(lr){2-7}
         $\alpha$ & \Dzero-meson~selections & Signal extraction  & Non-prompt \Dzero & Trk. eff. & Unfolding & Total  \\
        \midrule
        1 & 1--6\% &  1--7\% &  1--9\% & 1--9\%  & 1--7\%  & 2--16\%   \\
        1.5 & 1--5\% &  1--2\% &  1--10\% & 1--6\%  & 2--5\%  & 3--14\%   \\
        2 & 2--8\% & 1--2\% & 1--10\% & 1--8\%  & 2--7\%  & 3--17\%   \\
        3 & 2--8\% & 1--4\% & 1--7\% & 1--9\%  & 1--8\% & 3--17\%  \\
        \bottomrule
    \end{tabular}
    \label{tab:uncertainties_d0}
\end{table}

The sources of systematic uncertainty and their estimation for the inclusive-jet angularities follow the same procedure as the published results in Ref.~\cite{Alice_ang_2022}.  The total systematic uncertainty ranges from 2\% to 6\% for inclusive jets with the leading-track \pt selection, and from 5\% to 13\% without the selection, across the $\lambda_{\alpha}$ intervals. In both cases, the dominant source of systematic uncertainty is the tracking efficiency, which contributes 2\% to 5\% with the \pt selection on the leading track and 4\% to 12\% without the selection, while the remaining sources are small.
 
\section{Results}
\label{sec:results}
The jet angularities $\lambda_{\alpha}$ are reported as self-normalized distributions, 
\begin{equation}
    \frac{1}{\sigma_{\rm jet}}\frac{{\rm d} \sigma}{{\rm d} \lambda_{\alpha}} = \frac{1}{N_{\rm jet}}\frac{{\rm d} N}{{\rm d} \lambda_{\alpha}},
\end{equation}
where $N_{\rm jet}$ is the number of charged-particle jets in a given event sample for a given $\pTchjet$, and $\sigma_{\rm jet}$ is the corresponding cross section.  

\subsection{${\mathbf{D^{0}}}$-tagged and inclusive jet angularities}
\begin{figure}[!htb]
\centering
   \includegraphics[scale=0.35]{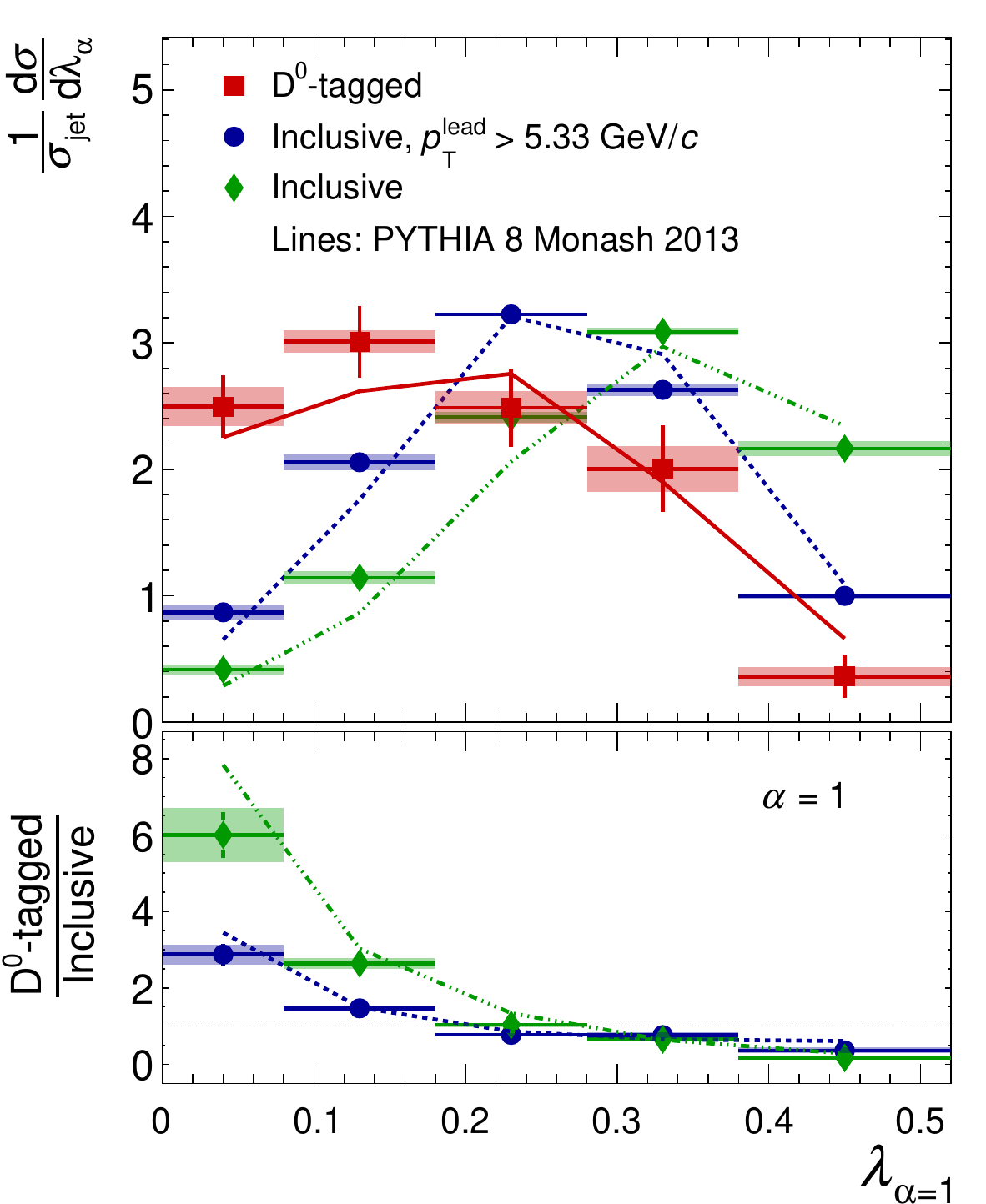}
    \includegraphics[scale=0.35]{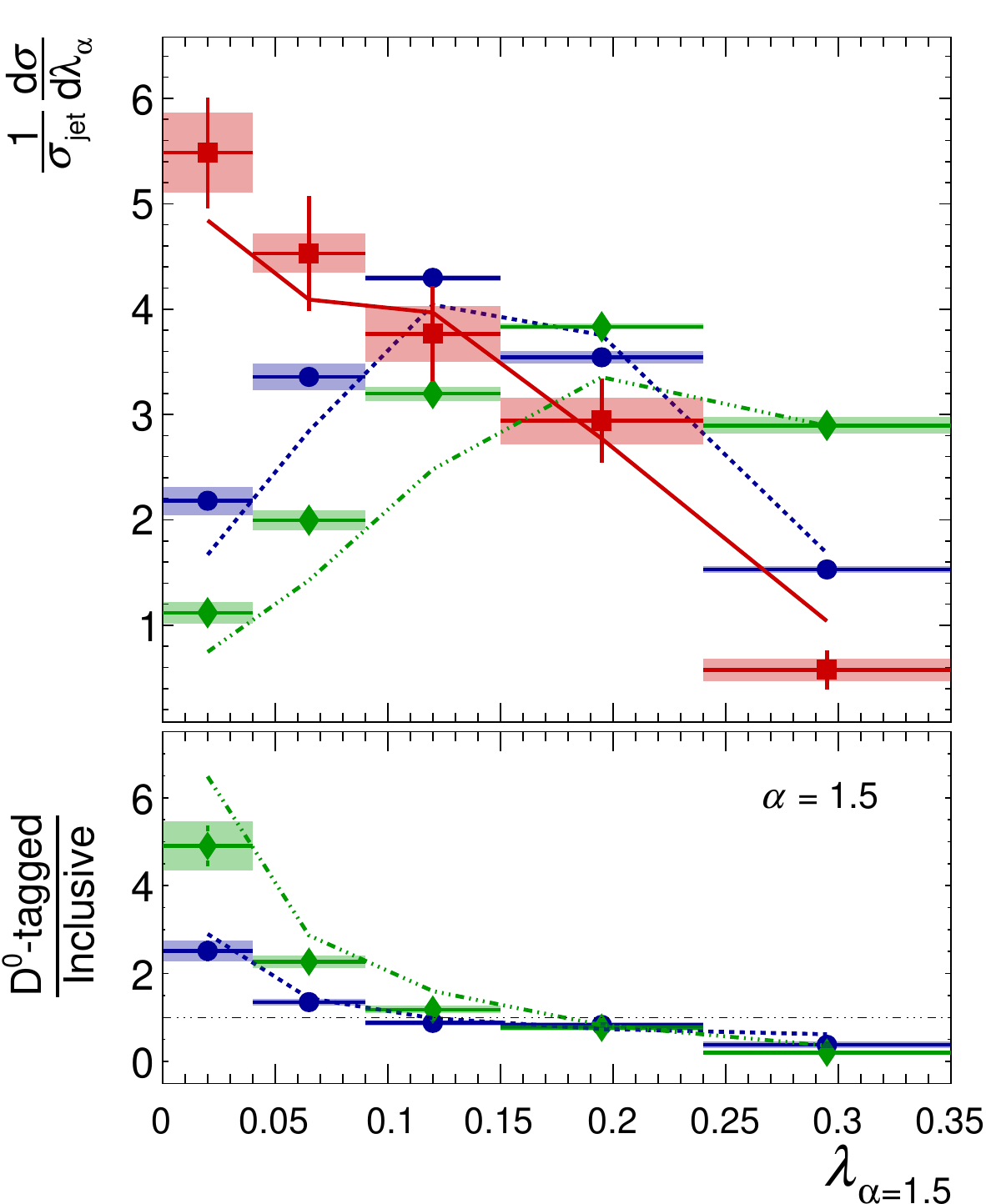}
     \includegraphics[scale=0.35]{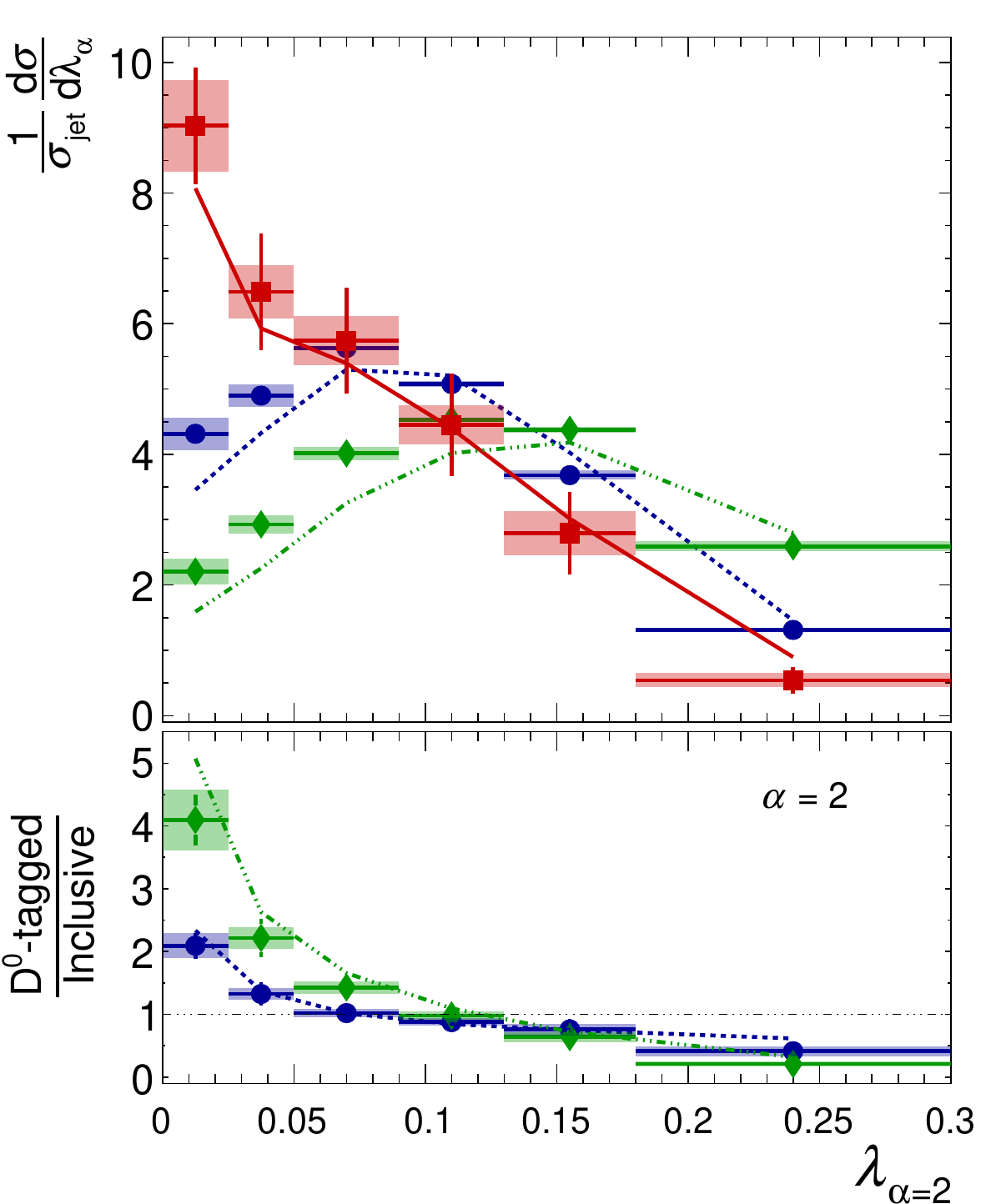}
    \includegraphics[scale=0.35]{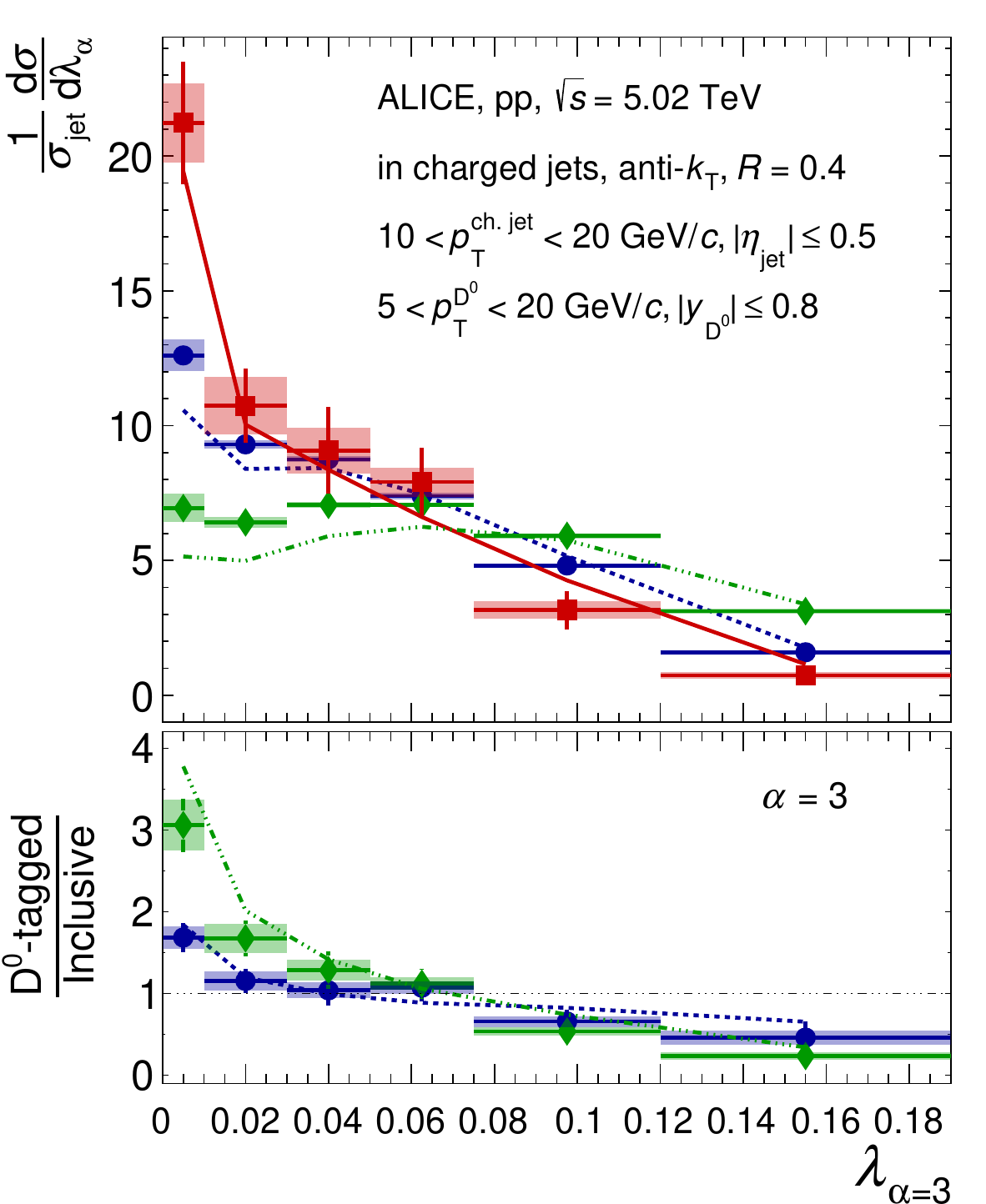}
    \caption{ The angularity distributions, measured by ALICE in pp collisions at $\sqrt{s}=5.02$ TeV, of \Dzero-tagged  jets (red) compared to inclusive charged-particle jets (blue and green) and to PYTHIA 8 simulations for $\alpha=1$, 1.5, 2, and 3 using $R=0.4$ in \jetptrange{10}{20}. The ratio of \Dzero-tagged to inclusive jet angularities is shown in the bottom panels for each value of $\alpha$. Note that the $y$-axis scale differs among the plots.}
\label{fig:alphascan_data}
\end{figure}

The fully corrected distributions of $\lambda_{\alpha}$ for \Dzero-tagged jets  and inclusive jets (a mixture of gluon- and quark-initiated jets) in the interval \jetptrange{10}{20} and \ptDrange{5}{20} are shown in Fig.~\ref{fig:alphascan_data}. Solid markers represent data, while the dashed lines correspond to PYTHIA 8 Monash 2013~\cite{Skands:2014pea} calculations. The $\lambda_{\alpha=1}$ distribution (upper left panel) for \Dzero-tagged jets peaks at lower values than inclusive jets with a leading-track \pt requirement; inclusive jets without a leading track requirement peak at an even larger value. This indicates that the transverse momentum carried by the jet fragments in \Dzero-tagged jets is more concentrated near the jet axis, whereas it is distributed to wider angles in inclusive jets. The peak shift is clear in the ratio panel, which shows an enhancement of \Dzero-tagged jet yields relative to inclusive jets at low angularities. The relative uncertainties are assumed to be uncorrelated between the \Dzero-tagged and inclusive jets. The enhancement is more pronounced when comparing to inclusive jets without the leading-track \pt selection. Recently, ALICE also compared energy–energy correlators (EECs) between \Dzero-tagged and inclusive jets, both with and without leading-track \pt selections~\cite{ALICE:EEC}. The leading-track selection impacts the shape of the angularities distributions more substantially than the EECs, where a shift in normalization is the dominant effect. This suggests that angularities are more sensitive to the emission angles of individual jet fragments.

The panels in Fig.~\ref{fig:alphascan_data}, from the upper left to the bottom right, illustrate the systematic variation of the jet angularity measurement with the weight parameter $\alpha$ from 1 to 1.5, 2, and 3. As the weight parameter $\alpha$ increases, the distributions become increasingly skewed toward smaller values of $\lambda_{\alpha}$, reflecting the fact that the angular term of Eq.~\ref{eq:angularity} $\theta_{i} \sim \Delta R_{i}/R$ is less than unity. With increasing $\alpha$, which gives higher weight to wide-angle radiation, the differences between the angularities of \Dzero-tagged jets and inclusive jets are reduced, as seen in the ratio panel at low angularity values.

\begin{figure}[!tb]
\centering
\includegraphics[width=0.8\textwidth]{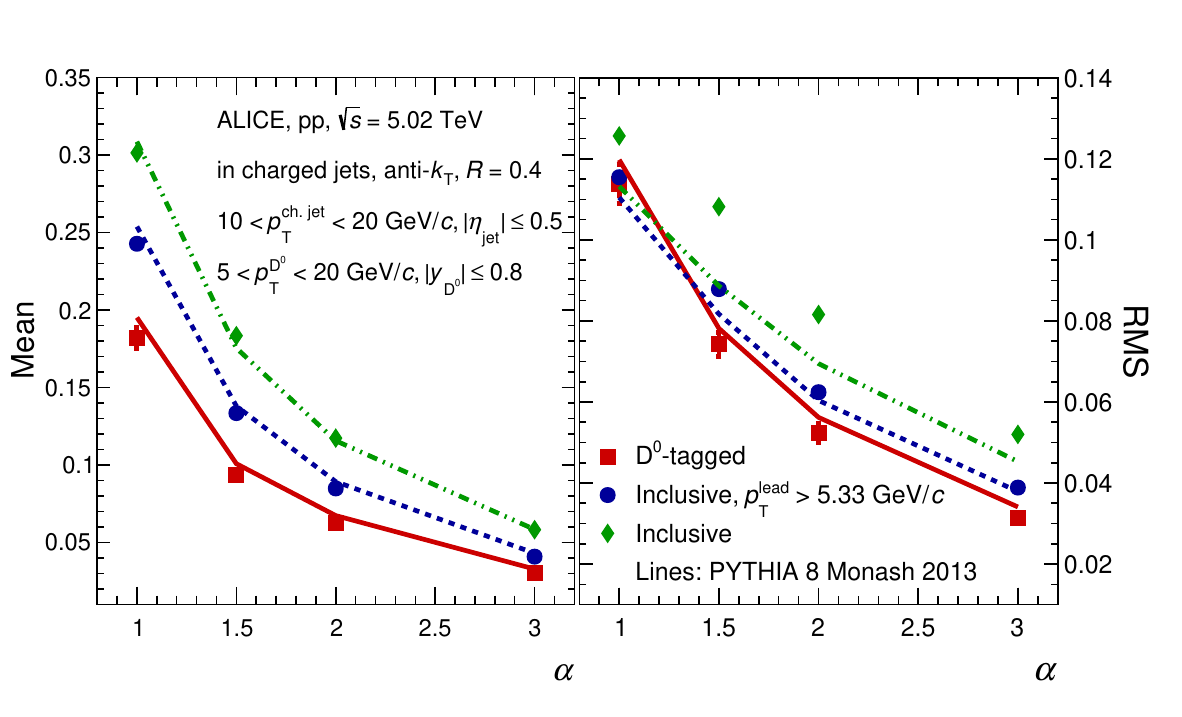}
\caption[]{The mean (left panel) and RMS (right panel) values extracted from the angularity distributions shown in Fig.~\ref{fig:alphascan_data} for \Dzero-tagged  jets and inclusive jets, with and without the leading-track \pt selection, as a function of $\alpha$ using $R=0.4$ in \jetptrange{10}{20}.}
\label{fig:mean_rms}

\end{figure}

The measurement is also compared to PYTHIA 8 in the figures. In the \jetptrange{10}{20} interval, the fraction of gluon-initiated jets in PYTHIA 8 is found to be 68\% with and 76\% without the leading-track \pt selection, indicating a slight bias of the selection toward quark-initiated jets. Overall, these predictions qualitatively describe both \Dzero-tagged  and inclusive jet-angularity distributions. Quantitatively, PYTHIA 8 provides a slightly better description of \Dzero-tagged jet angularities than those of inclusive jets, with the difference becoming more pronounced in the case without a leading-track \pt selection. The comparison to the ratio of \Dzero-tagged to inclusive jet angularities in the bottom panels shows good agreement (except for the first point), suggesting that PYTHIA 8 reasonably captures the relative fragmentation modifications. The observed trend is quantified in Fig.~\ref{fig:mean_rms}, which shows the shift in the peak positions (mean) and the root mean square (RMS) of the angularity distributions as a function of $\alpha$, and compares them to PYTHIA 8 predictions. The values are extracted from the distribution shown in Fig.~\ref{fig:alphascan_data}. Solid markers represent data, while the dashed lines correspond to PYTHIA 8 predictions. The angularity distributions for \Dzero-tagged jets are narrower and peak at smaller values of $\lambda_{\alpha}$ compared to inclusive jets with the leading-track $p_{\mathrm{T}}$ selection, and even more so when compared to inclusive jets without this selection. PYTHIA 8 shows excellent agreement with the mean values of the distributions, although significant deviations are observed in the RMS values for inclusive jets with no leading-track selection. This may indicate that the presence of an additional hard scale associated with the quark mass, in addition to the hard-scattering scale, leads to a better description of the distribution.

\begin{figure}[!tb]
\centering
   \includegraphics[scale=0.7]{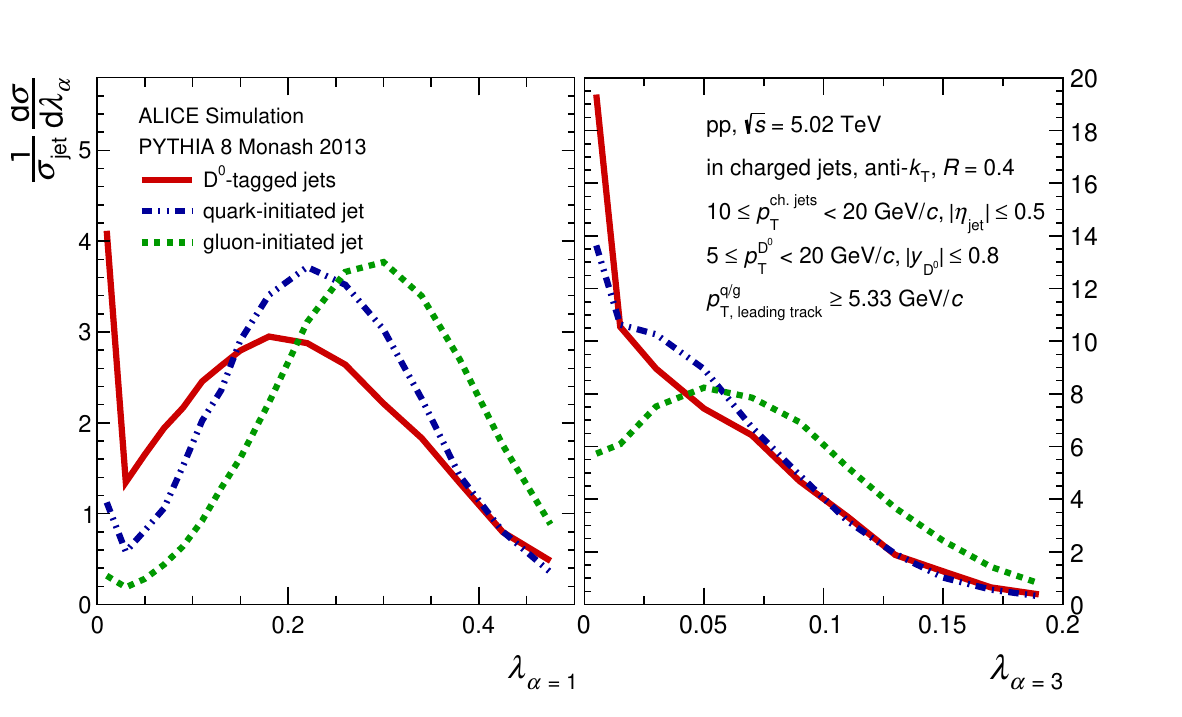}
    \caption{Comparison of jet angularity distributions for D$^{0}$-tagged (red), quark-initiated (blue), and gluon-initiated (green) jets from PYTHIA 8 Monash simulations with $R=0.4$ in the $10 < \pTchjet < 20$~\GeVc~range, shown for $\lambda_{\alpha= 1}$ (left) and $\lambda_{\alpha= 3}$ (right).}
\label{fig:PYTHIA8_angularity}
\end{figure}
The modification of $\lambda_{\alpha}$ distributions of \Dzero-tagged jets relative to inclusive jets, which consist of a mixture of quark- and gluon-initiated jets, is further investigated using PYTHIA~8 simulations. The angularities predicted by PYTHIA~8 for jets initiated by partons of different flavors are shown with finer binning than the data, represented by lines in Fig.~\ref{fig:PYTHIA8_angularity}. Such comparisons enable isolation of quark-mass effects from color-charge effects. The left panel shows the comparison for $\alpha=1$ where the jet core is given a higher weight, which emphasizes collinear radiation. In this case, quark-initiated jets exhibit lower angularities than gluon-initiated jets. \Dzero-tagged jets, enriched in charm quarks, have even lower values, demonstrating a clear sensitivity of $\lambda_{\alpha=1}$ to parton-mass effects such as the dead-cone and the harder fragmentation of heavy quarks. The \Dzero meson, included in the angularity calculation, typically carries a large fraction of the jet \pt~\cite{2023} and is close to the jet axis~\cite{Acharya_2025}, which shifts the distribution towards smaller angularity. This flavor hierarchy as observed in PYTHIA 8, is consistent with the data. 

The right panel shows $\lambda_{\alpha=3}$ angularities, where the wide-angle radiation is emphasized. Here, \Dzero-tagged and light-quark-initiated jets are similar, yet remain distinct from gluon-initiated jets, showing a clear sensitivity to color-charge effects. A similar trend is observed in the data when comparing to inclusive jets with and without a leading-track selection, as shown in Fig.~\ref{fig:alphascan_data}. The reduction in the differences between \Dzero-tagged and inclusive jets, visible in the ratio panel, is driven by the increased sensitivity to wide-angle radiation, indicating that the impact of the dead-cone effect decreases at larger $\alpha$, thereby enhancing sensitivity to color-charge (Casimir) differences between quark- and gluon-initiated jets.

\subsection{Investigating hadronization using MC model comparisons}
\begin{figure}[!tb]
\centering
\includegraphics[width=0.45\textwidth]{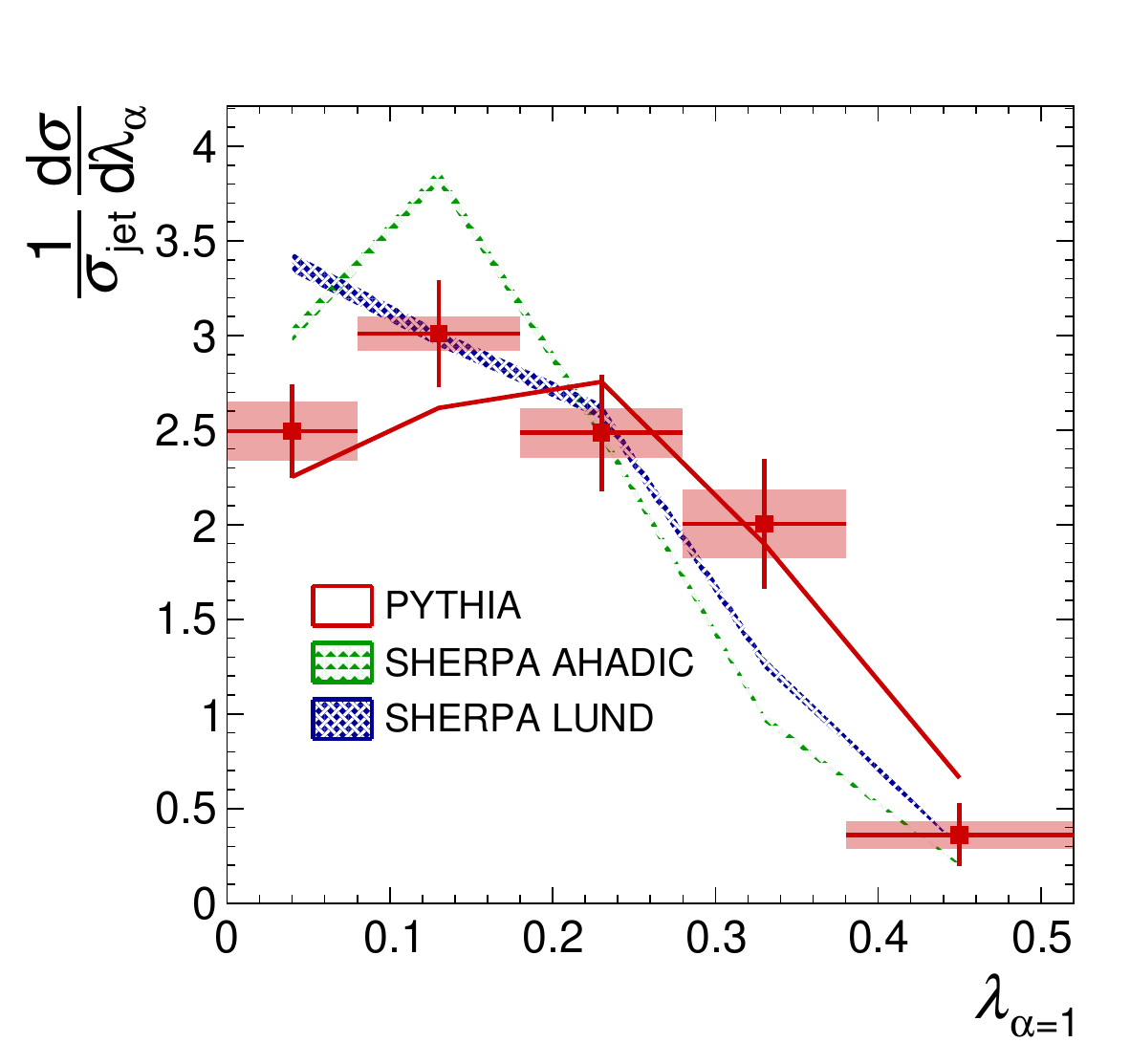}
\includegraphics[width=0.45\textwidth]{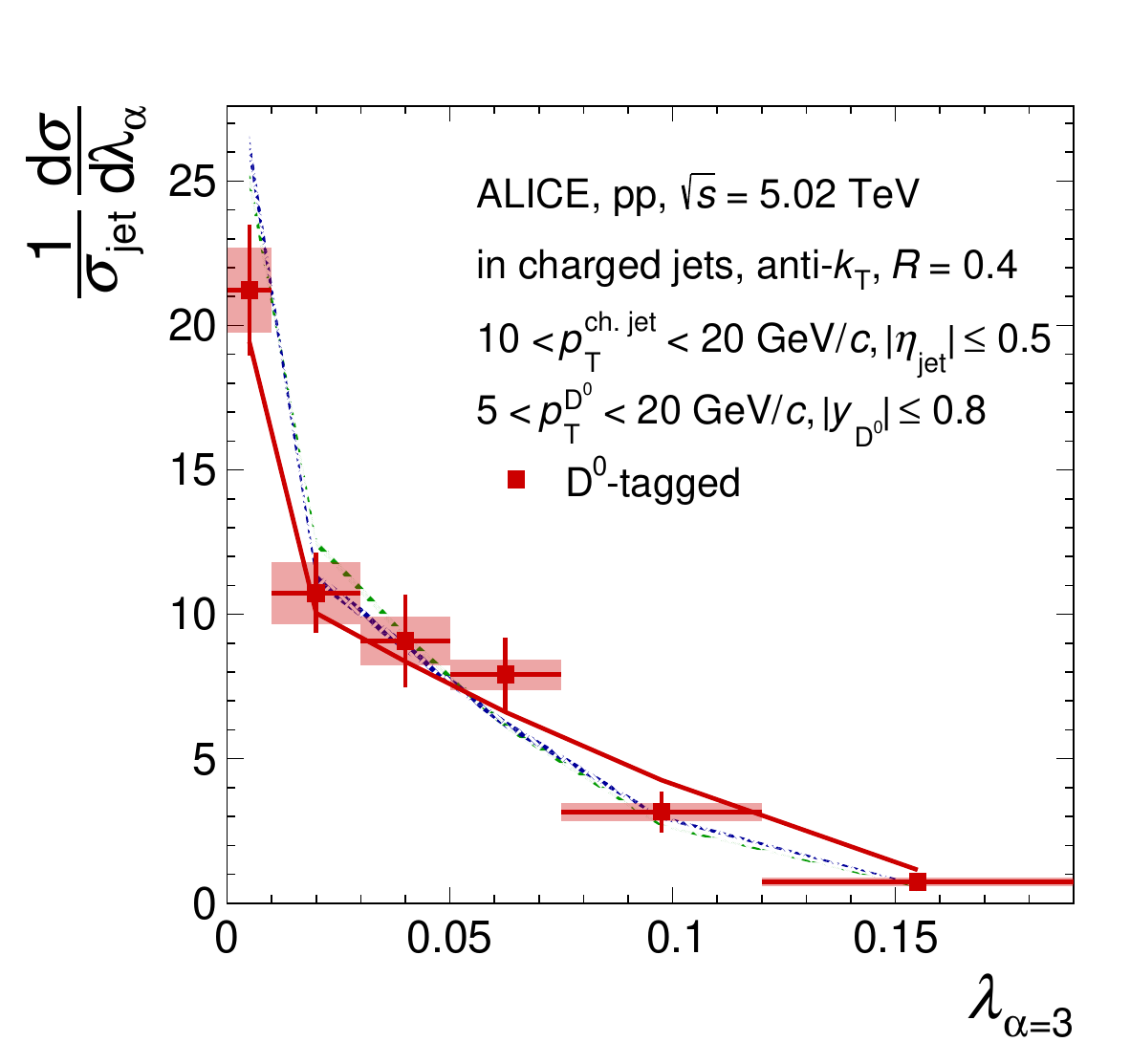}
\caption[]{The \Dzero-tagged jet-angularity distributions, measured by ALICE in pp collisions at $\sqrt{s}=5.02$ TeV (red markers), compared to PYTHIA 8 Monash (red line), SHERPA AHADIC (green line), and SHERPA LUND (blue line) using $R=0.4$ in \jetptrange{10}{20}, for $\lambda_{\alpha= 1}$ (left) and $\lambda_{\alpha= 3}$ (right).}
\label{fig:MC_models}
\end{figure}

In order to investigate the role of non-perturbative effects such as hadronization, the \Dzero-tagged jet angularities are compared in Fig.~\ref{fig:MC_models} with MC predictions obtained using different parton-shower and hadronization models. The left panel shows the results for $\lambda_{\alpha=1}$, while the right panel corresponds to $\lambda_{\alpha=3}$. Among the tested generators, PYTHIA~8 Monash, which employs the Lund string model, provides the best description of the \Dzero-tagged jet-angularity distributions.

Calculations from SHERPA 2.2.15~\cite{10.21468/SciPostPhys.7.3.034},
which allows the parton-shower and hadronization tunes to be modified independently, are
also shown. In this setup, the parton shower is kept fixed while only the hadronization model is varied. SHERPA-LUND, which implements the Lund string hadronization model, and SHERPA-AHADIC, which uses a cluster-hadronization model, both show some deviations near the peak of the distributions for $\lambda_{\alpha=1}$, while they yield similar results for $\lambda_{\alpha=3}$.

\section{Conclusions}
\label{sec:conclusions}
This paper presents the first measurement of \Dzero-tagged jets (charm-enriched) angularities in \pp~collisions at $\sqrt{s} = 5.02$~\TeV. Jet angularities are powerful substructure observables that differ between quark- and gluon-initiated jets, and reflect the flavor dependence of parton fragmentation. The charm-enriched angularities of \Dzero-tagged jets are compared with those of inclusive (gluon-dominated) jets in the \mbox{$10 < \pTchjet < 20$~\GeVc} interval, where charm-quark mass effects are most significant. For $\alpha = 1$, which emphasizes the contribution from collinear radiation, \Dzero-tagged jets show that the transverse momentum carried by the jet fragments is more concentrated near the jet axis than that of inclusive jets. This observation is consistent with the QCD dead-cone effect. For larger values of the weight parameter $\alpha$, corresponding to enhanced sensitivity to wide-angle radiation, \Dzero-tagged and inclusive jets are more similar, reflecting the reduced sensitivity to parton mass and increased sensitivity to color-charge effects. Furthermore, the comparison between jet angularities and energy-energy correlators suggests that jet angularities are more sensitive to the emission angles of individual jet fragments, as evidenced by their shape dependence on the leading-track selection.

The experimental results are compared with PYTHIA 8 simulations, which qualitatively describe the peak of the angularity distributions for both \Dzero-tagged and inclusive jets, but show deviations in the shape of the inclusive-jet distributions. These results provide constraints for refining the description of heavy-quark jets in QCD calculations and MC simulations. To investigate the role of hadronization, the measurements are further compared with SHERPA-AHADIC and SHERPA-LUND simulations implementing different hadronization tunes. Both tunes exhibit noticeable deviations near the peak of the $\lambda_{\alpha=1}$ distribution, while yielding similar results for $\lambda_{\alpha=3}$.

Future studies extending this analysis to higher jet \pt and to beauty-tagged jets will offer further insights into the flavor dependence of jet substructure, tracing the transition from a region dominated by quark-mass effects, such as the dead-cone effect, to higher \pt where these effects diminish and sensitivity to color-charge effects becomes more pronounced. These measurements also provide an essential vacuum baseline for future studies of jet angularities in heavy-ion collisions, enabling a deeper understanding of how QCD dynamics is modified in the presence of the quark–gluon plasma.

\newenvironment{acknowledgement}{\relax}{\relax}
\begin{acknowledgement}
\section*{Acknowledgments}

The ALICE Collaboration would like to thank all its engineers and technicians for their invaluable contributions to the construction of the experiment and the CERN accelerator teams for the outstanding performance of the LHC complex.
The ALICE Collaboration gratefully acknowledges the resources and support provided by all Grid centres and the Worldwide LHC Computing Grid (WLCG) collaboration.
The ALICE Collaboration acknowledges the following funding agencies for their support in building and running the ALICE detector:
A. I. Alikhanyan National Science Laboratory (Yerevan Physics Institute) Foundation (ANSL), State Committee of Science and World Federation of Scientists (WFS), Armenia;
Austrian Academy of Sciences, Austrian Science Fund (FWF): [M 2467-N36] and Nationalstiftung f\"{u}r Forschung, Technologie und Entwicklung, Austria;
Ministry of Communications and High Technologies, National Nuclear Research Center, Azerbaijan;
Rede Nacional de Física de Altas Energias (Renafae), Financiadora de Estudos e Projetos (Finep), Funda\c{c}\~{a}o de Amparo \`{a} Pesquisa do Estado de S\~{a}o Paulo (FAPESP) and The Sao Paulo Research Foundation  (FAPESP), Brazil;
Bulgarian Ministry of Education and Science, within the National Roadmap for Research Infrastructures 2020-2027 (object CERN), Bulgaria;
Ministry of Education of China (MOEC) , Ministry of Science \& Technology of China (MSTC) and National Natural Science Foundation of China (NSFC), China;
Ministry of Science and Education and Croatian Science Foundation, Croatia;
Centro de Aplicaciones Tecnol\'{o}gicas y Desarrollo Nuclear (CEADEN), Cubaenerg\'{\i}a, Cuba;
Ministry of Education, Youth and Sports of the Czech Republic, Czech Republic;
The Danish Council for Independent Research | Natural Sciences, the VILLUM FONDEN and Danish National Research Foundation (DNRF), Denmark;
Helsinki Institute of Physics (HIP), Finland;
Commissariat \`{a} l'Energie Atomique (CEA) and Institut National de Physique Nucl\'{e}aire et de Physique des Particules (IN2P3) and Centre National de la Recherche Scientifique (CNRS), France;
Bundesministerium f\"{u}r Forschung, Technologie und Raumfahrt (BMFTR) and GSI Helmholtzzentrum f\"{u}r Schwerionenforschung GmbH, Germany;
National Research, Development and Innovation Office, Hungary;
Department of Atomic Energy Government of India (DAE), Department of Science and Technology, Government of India (DST), University Grants Commission, Government of India (UGC) and Council of Scientific and Industrial Research (CSIR), India;
National Research and Innovation Agency - BRIN, Indonesia;
Istituto Nazionale di Fisica Nucleare (INFN), Italy;
Japanese Ministry of Education, Culture, Sports, Science and Technology (MEXT) and Japan Society for the Promotion of Science (JSPS) KAKENHI, Japan;
Consejo Nacional de Ciencia (CONACYT) y Tecnolog\'{i}a, through Fondo de Cooperaci\'{o}n Internacional en Ciencia y Tecnolog\'{i}a (FONCICYT) and Direcci\'{o}n General de Asuntos del Personal Academico (DGAPA), Mexico;
Nederlandse Organisatie voor Wetenschappelijk Onderzoek (NWO), Netherlands;
The Research Council of Norway, Norway;
Pontificia Universidad Cat\'{o}lica del Per\'{u}, Peru;
Ministry of Science and Higher Education, National Science Centre and WUT ID-UB, Poland;
Korea Institute of Science and Technology Information and National Research Foundation of Korea (NRF), Republic of Korea;
Ministry of Education and Scientific Research, Institute of Atomic Physics, Ministry of Research and Innovation and Institute of Atomic Physics and Universitatea Nationala de Stiinta si Tehnologie Politehnica Bucuresti, Romania;
Ministerstvo skolstva, vyskumu, vyvoja a mladeze SR, Slovakia;
National Research Foundation of South Africa, South Africa;
Swedish Research Council (VR) and Knut \& Alice Wallenberg Foundation (KAW), Sweden;
European Organization for Nuclear Research, Switzerland;
Suranaree University of Technology (SUT), National Science and Technology Development Agency (NSTDA) and National Science, Research and Innovation Fund (NSRF via PMU-B B05F650021), Thailand;
Turkish Energy, Nuclear and Mineral Research Agency (TENMAK), Turkey;
National Academy of  Sciences of Ukraine, Ukraine;
Science and Technology Facilities Council (STFC), United Kingdom;
National Science Foundation of the United States of America (NSF) and United States Department of Energy, Office of Nuclear Physics (DOE NP), United States of America.
In addition, individual groups or members have received support from:
Czech Science Foundation (grant no. 23-07499S), Czech Republic;
FORTE project, reg.\ no.\ CZ.02.01.01/00/22\_008/0004632, Czech Republic, co-funded by the European Union, Czech Republic;
European Research Council (grant nos. 101220549, 950692), European Union;
Deutsche Forschungs Gemeinschaft (DFG, German Research Foundation) ``Neutrinos and Dark Matter in Astro- and Particle Physics'' (grant no. SFB 1258), Germany;
CONVECS project, CUP C97H23001700002 FESR 2021-2027 program, Italy.

\end{acknowledgement}

\bibliographystyle{utphys}   
\bibliography{main}

\newpage
\appendix
\section{The ALICE Collaboration}
\label{app:collab}
\begin{flushleft} 
\small

D.A.H.~Abdallah\,\orcidlink{0000-0003-4768-2718}\,$^{\rm 134}$, 
I.J.~Abualrob\,\orcidlink{0009-0005-3519-5631}\,$^{\rm 112}$, 
S.~Acharya\,\orcidlink{0000-0002-9213-5329}\,$^{\rm 49}$, 
K.~Agarwal\,\orcidlink{0000-0001-5781-3393}\,$^{\rm II,}$$^{\rm 23}$, 
G.~Aglieri Rinella\,\orcidlink{0000-0002-9611-3696}\,$^{\rm 32}$, 
L.~Aglietta\,\orcidlink{0009-0003-0763-6802}\,$^{\rm 24}$, 
N.~Agrawal\,\orcidlink{0000-0003-0348-9836}\,$^{\rm 25}$, 
Z.~Ahammed\,\orcidlink{0000-0001-5241-7412}\,$^{\rm 132}$, 
S.~Ahmad\,\orcidlink{0000-0003-0497-5705}\,$^{\rm 15}$, 
Z.~Akbar$^{\rm 79}$
% ...

\section*{Affiliation Notes}

$^{\rm I}$ Deceased\\
$^{\rm II}$ Also at: INFN Trieste, Trieste, Italy\\
$^{\rm III}$ Also at: Fondazione Bruno Kessler (FBK), Trento, Italy\\
$^{\rm IV}$ Also at: Czech Technical University in Prague, Prague, Czech Republic\\
$^{\rm V}$ Also at: Instituto de Fisica da Universidade de Sao Paulo\\
$^{\rm VI}$ Also at: Dipartimento DET del Politecnico di Torino, Turin, Italy\\
$^{\rm VII}$ Also at: University College of Dublin, Dublin, Ireland\\
$^{\rm VIII}$ Also at: Department of Applied Physics, Aligarh Muslim University, Aligarh, India\\
$^{\rm IX}$ Also at: Institute of Theoretical Physics, University of Wroclaw, Wroclaw, Poland\\
$^{\rm X}$ Also at: Facultad de Ciencias, Universidad Nacional Aut\'{o}noma de M\'{e}xico, Mexico City, Mexico\\

\section*{Collaboration Institutes}

$^{1}$ A.I. Alikhanyan National Science Laboratory (Yerevan Physics Institute) Foundation, Yerevan, Armenia\\
$^{2}$ AGH University of Krakow, Cracow, Poland\\
$^{3}$ Bogolyubov Institute for Theoretical Physics, National Academy of Sciences of Ukraine, Kyiv, Ukraine\\
$^{4}$ Bose Institute, Department of Physics  and Centre for Astroparticle Physics and Space Science (CAPSS), Kolkata, India\\
$^{5}$ California Polytechnic State University, San Luis Obispo, California, United States\\
$^{6}$ Central China Normal University, Wuhan, China\\
$^{7}$ Centro de Aplicaciones Tecnol\'{o}gicas y Desarrollo Nuclear (CEADEN), Havana, Cuba\\
$^{8}$ Centro de Investigaci\'{o}n y de Estudios Avanzados (CINVESTAV), Mexico City and M\'{e}rida, Mexico\\
$^{9}$ Chicago State University, Chicago, Illinois, United States\\
$^{10}$ China Nuclear Data Center, China Institute of Atomic Energy, Beijing, China\\
$^{11}$ China University of Geosciences, Wuhan, China\\
$^{12}$ Chungbuk National University, Cheongju, Republic of Korea\\
$^{13}$ Comenius University Bratislava, Faculty of Mathematics, Physics and Informatics, Bratislava, Slovak Republic\\
$^{14}$ Creighton University, Omaha, Nebraska, United States\\
$^{15}$ Department of Physics, Aligarh Muslim University, Aligarh, India\\
$^{16}$ Department of Physics, Pusan National University, Pusan, Republic of Korea\\
$^{17}$ Department of Physics, Sejong University, Seoul, Republic of Korea\\
$^{18}$ Department of Physics, University of California, Berkeley, California, United States\\
$^{19}$ Department of Physics, University of Oslo, Oslo, Norway\\
$^{20}$ Department of Physics and Technology, University of Bergen, Bergen, Norway\\
$^{21}$ Dipartimento di Fisica, Universit\`{a} di Pavia, Pavia, Italy\\
$^{22}$ Dipartimento di Fisica dell'Universit\`{a} and Sezione INFN, Cagliari, Italy\\
$^{23}$ Dipartimento di Fisica dell'Universit\`{a} and Sezione INFN, Trieste, Italy\\
$^{24}$ Dipartimento di Fisica dell'Universit\`{a} and Sezione INFN, Turin, Italy\\
$^{25}$ Dipartimento di Fisica e Astronomia dell'Universit\`{a} and Sezione INFN, Bologna, Italy\\
$^{26}$ Dipartimento di Fisica e Astronomia dell'Universit\`{a} and Sezione INFN, Catania, Italy\\
$^{27}$ Dipartimento di Fisica e Astronomia dell'Universit\`{a} and Sezione INFN, Padova, Italy\\
$^{28}$ Dipartimento di Fisica `E.R.~Caianiello' dell'Universit\`{a} and Gruppo Collegato INFN, Salerno, Italy\\
$^{29}$ Dipartimento DISAT del Politecnico and Sezione INFN, Turin, Italy\\
$^{30}$ Dipartimento di Scienze MIFT, Universit\`{a} di Messina, Messina, Italy\\
$^{31}$ Dipartimento Interateneo di Fisica `M.~Merlin' and Sezione INFN, Bari, Italy\\
$^{32}$ European Organization for Nuclear Research (CERN), Geneva, Switzerland\\
$^{33}$ Faculty of Electrical Engineering, Mechanical Engineering and Naval Architecture, University of Split, Split, Croatia\\
$^{34}$ Faculty of Nuclear Sciences and Physical Engineering, Czech Technical University in Prague, Prague, Czech Republic\\
$^{35}$ Faculty of Physics, Sofia University, Sofia, Bulgaria\\
$^{36}$ Faculty of Science, P.J.~\v{S}af\'{a}rik University, Ko\v{s}ice, Slovak Republic\\
$^{37}$ Faculty of Technology, Environmental and Social Sciences, Bergen, Norway\\
$^{38}$ Frankfurt Institute for Advanced Studies, Johann Wolfgang Goethe-Universit\"{a}t Frankfurt, Frankfurt, Germany\\
$^{39}$ Fudan University, Shanghai, China\\
$^{40}$ Gauhati University, Department of Physics, Guwahati, India\\
$^{41}$ Helmholtz-Institut f\"{u}r Strahlen- und Kernphysik, Rheinische Friedrich-Wilhelms-Universit\"{a}t Bonn, Bonn, Germany\\
$^{42}$ Helsinki Institute of Physics (HIP), Helsinki, Finland\\
$^{43}$ High Energy Physics Group,  Universidad Aut\'{o}noma de Puebla, Puebla, Mexico\\
$^{44}$ Horia Hulubei National Institute of Physics and Nuclear Engineering, Bucharest, Romania\\
$^{45}$ HUN-REN Wigner Research Centre for Physics, Budapest, Hungary\\
$^{46}$ Indian Institute of Technology Bombay (IIT), Mumbai, India\\
$^{47}$ Indian Institute of Technology Indore, Indore, India\\
$^{48}$ INFN, Laboratori Nazionali di Frascati, Frascati, Italy\\
$^{49}$ INFN, Sezione di Bari, Bari, Italy\\
$^{50}$ INFN, Sezione di Bologna, Bologna, Italy\\
$^{51}$ INFN, Sezione di Cagliari, Cagliari, Italy\\
$^{52}$ INFN, Sezione di Catania, Catania, Italy\\
$^{53}$ INFN, Sezione di Padova, Padova, Italy\\
$^{54}$ INFN, Sezione di Pavia, Pavia, Italy\\
$^{55}$ INFN, Sezione di Torino, Turin, Italy\\
$^{56}$ INFN, Sezione di Trieste, Trieste, Italy\\
$^{57}$ Inha University, Incheon, Republic of Korea\\
$^{58}$ Institute for Gravitational and Subatomic Physics (GRASP), Utrecht University/Nikhef, Utrecht, Netherlands\\
$^{59}$ Institute of Experimental Physics, Slovak Academy of Sciences, Ko\v{s}ice, Slovak Republic\\
$^{60}$ Institute of Physics, Homi Bhabha National Institute, Bhubaneswar, India\\
$^{61}$ Institute of Physics of the Czech Academy of Sciences, Prague, Czech Republic\\
$^{62}$ Institute of Space Science (ISS), Bucharest, Romania\\
$^{63}$ Institut f\"{u}r Kernphysik, Johann Wolfgang Goethe-Universit\"{a}t Frankfurt, Frankfurt, Germany\\
$^{64}$ Instituto de Ciencias Nucleares, Universidad Nacional Aut\'{o}noma de M\'{e}xico, Mexico City, Mexico\\
$^{65}$ Instituto de F\'{i}sica, Universidade Federal do Rio Grande do Sul (UFRGS), Porto Alegre, Brazil\\
$^{66}$ Instituto de F\'{\i}sica, Universidad Nacional Aut\'{o}noma de M\'{e}xico, Mexico City, Mexico\\
$^{67}$ iThemba LABS, National Research Foundation, Somerset West, South Africa\\
$^{68}$ Jeonbuk National University, Jeonju, Republic of Korea\\
$^{69}$ Korea Institute of Science and Technology Information, Daejeon, Republic of Korea\\
$^{70}$ Laboratoire de Physique Subatomique et de Cosmologie, Universit\'{e} Grenoble-Alpes, CNRS-IN2P3, Grenoble, France\\
$^{71}$ Lawrence Berkeley National Laboratory, Berkeley, California, United States\\
$^{72}$ Lund University Department of Physics, Division of Particle Physics, Lund, Sweden\\
$^{73}$ Marietta Blau Institute, Vienna, Austria\\
$^{74}$ Nagasaki Institute of Applied Science, Nagasaki, Japan\\
$^{75}$ Nara Women{'}s University (NWU), Nara, Japan\\
$^{76}$ National Centre for Nuclear Research, Warsaw, Poland\\
$^{77}$ National Institute of Science Education and Research, Homi Bhabha National Institute, Jatni, India\\
$^{78}$ National Nuclear Research Center, Baku, Azerbaijan\\
$^{79}$ National Research and Innovation Agency - BRIN, Jakarta, Indonesia\\
$^{80}$ Niels Bohr Institute, University of Copenhagen, Copenhagen, Denmark\\
$^{81}$ Nikhef, National institute for subatomic physics, Amsterdam, Netherlands\\
$^{82}$ Nuclear Physics Group, STFC Daresbury Laboratory, Daresbury, United Kingdom\\
$^{83}$ Nuclear Physics Institute of the Czech Academy of Sciences, Husinec-\v{R}e\v{z}, Czech Republic\\
$^{84}$ Oak Ridge National Laboratory, Oak Ridge, Tennessee, United States\\
$^{85}$ Ohio State University, Columbus, Ohio, United States\\
$^{86}$ Physics department, Faculty of science, University of Zagreb, Zagreb, Croatia\\
$^{87}$ Physics Department, Panjab University, Chandigarh, India\\
$^{88}$ Physics Department, University of Jammu, Jammu, India\\
$^{89}$ Physics Program and International Institute for Sustainability with Knotted Chiral Meta Matter (WPI-SKCM$^{2}$), Hiroshima University, Hiroshima, Japan\\
$^{90}$ Physikalisches Institut, Eberhard-Karls-Universit\"{a}t T\"{u}bingen, T\"{u}bingen, Germany\\
$^{91}$ Physikalisches Institut, Ruprecht-Karls-Universit\"{a}t Heidelberg, Heidelberg, Germany\\
$^{92}$ Physik Department, Technische Universit\"{a}t M\"{u}nchen, Munich, Germany\\
$^{93}$ Politecnico di Bari and Sezione INFN, Bari, Italy\\
$^{94}$ Research Division and ExtreMe Matter Institute EMMI, GSI Helmholtzzentrum f\"ur Schwerionenforschung GmbH, Darmstadt, Germany\\
$^{95}$ Saga University, Saga, Japan\\
$^{96}$ Saha Institute of Nuclear Physics, Homi Bhabha National Institute, Kolkata, India\\
$^{97}$ School of Physics and Astronomy, University of Birmingham, Birmingham, United Kingdom\\
$^{98}$ Secci\'{o}n F\'{\i}sica, Departamento de Ciencias, Pontificia Universidad Cat\'{o}lica del Per\'{u}, Lima, Peru\\
$^{99}$ SUBATECH, IMT Atlantique, Nantes Universit\'{e}, CNRS-IN2P3, Nantes, France\\
$^{100}$ Sungkyunkwan University, Suwon City, Republic of Korea\\
$^{101}$ Suranaree University of Technology, Nakhon Ratchasima, Thailand\\
$^{102}$ Technical University of Ko\v{s}ice, Ko\v{s}ice, Slovak Republic\\
$^{103}$ The Henryk Niewodniczanski Institute of Nuclear Physics, Polish Academy of Sciences, Cracow, Poland\\
$^{104}$ The University of Texas at Austin, Austin, Texas, United States\\
$^{105}$ Universidad Aut\'{o}noma de Sinaloa, Culiac\'{a}n, Mexico\\
$^{106}$ Universidade de S\~{a}o Paulo (USP), S\~{a}o Paulo, Brazil\\
$^{107}$ Universidade Estadual de Campinas (UNICAMP), Campinas, Brazil\\
$^{108}$ Universidade Federal do ABC, Santo Andre, Brazil\\
$^{109}$ Universitatea Nationala de Stiinta si Tehnologie Politehnica Bucuresti, Bucharest, Romania\\
$^{110}$ University of Cape Town, Cape Town, South Africa\\
$^{111}$ University of Derby, Derby, United Kingdom\\
$^{112}$ University of Houston, Houston, Texas, United States\\
$^{113}$ University of Jyv\"{a}skyl\"{a}, Jyv\"{a}skyl\"{a}, Finland\\
$^{114}$ University of Kansas, Lawrence, Kansas, United States\\
$^{115}$ University of Liverpool, Liverpool, United Kingdom\\
$^{116}$ University of Science and Technology of China, Hefei, China\\
$^{117}$ University of Silesia in Katowice, Katowice, Poland\\
$^{118}$ University of South-Eastern Norway, Kongsberg, Norway\\
$^{119}$ University of Tennessee, Knoxville, Tennessee, United States\\
$^{120}$ University of the Witwatersrand, Johannesburg, South Africa\\
$^{121}$ University of Tokyo, Tokyo, Japan\\
$^{122}$ University of Tsukuba, Tsukuba, Japan\\
$^{123}$ Universit\"{a}t M\"{u}nster, Institut f\"{u}r Kernphysik, M\"{u}nster, Germany\\
$^{124}$ Universit\'{e} Clermont Auvergne, CNRS/IN2P3, LPC, Clermont-Ferrand, France\\
$^{125}$ Universit\'{e} de Lyon, CNRS/IN2P3, Institut de Physique des 2 Infinis de Lyon, Lyon, France\\
$^{126}$ Universit\'{e} de Strasbourg, CNRS, IPHC UMR 7178, F-67000 Strasbourg, France, Strasbourg, France\\
$^{127}$ Universit\'{e} Paris-Saclay, Centre d'Etudes de Saclay (CEA), IRFU, D\'{e}partment de Physique Nucl\'{e}aire (DPhN), Saclay, France\\
$^{128}$ Universit\'{e}  Paris-Saclay, CNRS/IN2P3, IJCLab, Orsay, France\\
$^{129}$ Universit\`{a} degli Studi di Foggia, Foggia, Italy\\
$^{130}$ Universit\`{a} del Piemonte Orientale, Vercelli, Italy\\
$^{131}$ Universit\`{a} di Brescia, Brescia, Italy\\
$^{132}$ Variable Energy Cyclotron Centre, Homi Bhabha National Institute, Kolkata, India\\
$^{133}$ Warsaw University of Technology, Warsaw, Poland\\
$^{134}$ Wayne State University, Detroit, Michigan, United States\\
$^{135}$ Yale University, New Haven, Connecticut, United States\\
$^{136}$ Yildiz Technical University, Istanbul, Turkey\\
$^{137}$ Yonsei University, Seoul, Republic of Korea\\
$^{138}$ Affiliated with an institute formerly covered by a cooperation agreement with CERN\\
$^{139}$ Affiliated with an international laboratory covered by a cooperation agreement with CERN.\\

\end{flushleft} 
\end{document}